\documentclass[twocolumn,prd,nofootinbib]{revtex4}
\usepackage{graphicx}
\usepackage{dcolumn}
\usepackage{amsfonts}
\usepackage{amssymb}
\usepackage{amsmath}

\def\be{\begin{equation}}
\def\ee{\end{equation}}
\def\ba{\begin{eqnarray}}
\def\ea{\end{eqnarray}}

\begin{document}

\title{Gravitational dynamics in a 2+1+1 decomposed spacetime along
nonorthogonal double foliations: Hamiltonian evolution and gauge fixing}
\author{Cec\'{\i}lia Gergely}
\affiliation{Institute of Physics, University of Szeged, D\'{o}m t\'{e}r 9, 6720 Szeged,
Hungary}
\author{Zolt\'{a}n Keresztes}
\affiliation{Institute of Physics, University of Szeged, D\'{o}m t\'{e}r 9, 6720 Szeged,
Hungary}
\author{L\'{a}szl\'{o} \'{A}. Gergely}
\affiliation{Institute of Physics, University of Szeged, D\'{o}m t\'{e}r 9, 6720 Szeged,
Hungary}
\date{\today }

\begin{abstract}
Motivated by situations with temporal evolution and spatial symmetries both
singled out, we develop a new 2+1+1 decomposition of spacetime, based on a
nonorthogonal double foliation. Time evolution proceeds along the leaves of
the spatial foliation. We identify the gravitational variables in the
velocity phase-space as the 2-metric (induced on the intersection $\Sigma
_{t\chi }$ of the hypersurfaces of the foliations), the 2+1 components of
the spatial shift vector, together with the extrinsic curvature, normal
fundamental form and normal fundamental scalar of $\Sigma _{t\chi }$, all
constructed with the normal to the temporal foliation. This work generalizes
a previous decomposition based on orthogonal foliations, a formalism lacking
one metric variable, now reintroduced. The new metric variable is related to
(i) the angle of a Lorentz-rotation between the nonorthogonal bases adapted
to the foliations, and (ii) to the vorticity of these basis vectors. As a
first application of the formalism, we work out the Hamiltonian dynamics of
general relativity in terms of the variables identified as canonical,
generalizing previous work. As a second application we present the
unambiguous gauge-fixing suitable to discuss the even sector scalar-type
perturbations of spherically symmetric and static spacetimes in generic
scalar-tensor gravitational theories, which has been obstructed in the
formalism of orthogonal double foliation.
\end{abstract}

\maketitle

\section{Introduction}

\label{intro}

The modern theory of gravitation, general relativity (GR) has been
successfully tested multiple times on the Solar Sytem scale. When confronted
with observations on both galactic scales and beyond, agreement with
predictions can however be reached only at the price of introducing dark
matter and dark energy, neither of them identified or detected by other
means than gravitational. Lacking indications on manifestations of these
forms of matter in the Standard Model interactions, they could be included
in the gravitational sector, either as geometric modifications arising from
possible higher-order dynamics or as an excess of fields representing
gravity beyond the metric tensor, possibly including scalars, vectors,
2-form fields or even a second metric. As a rule, the physical metric
couples to these in a nonminimal way, opposed to dark matter/dark energy
models, which are coupled minimally. The simplest such model, of a single
scalar field complementing the metric has been studied extensively, both
from the desire to explain dark matter / dark energy or in order to study
inflation.

The most generic single scalar-tensor model described by second order
differential equations (hence avoiding Ostrogradski instabilities) for both
the metric and the scalar field has been proposed by Horndeski \cite%
{Horndeski} and rediscovered in a modern context in connection with
generalized galileons \cite{Deffayet}.

While allowing for higher order than two, certain beyond Horndeski models
could guarantee that the propagating degrees of freedom (d.o.f.) still evolve according to a
second order dynamics. Indeed, an effective field theory (EFT) of
cosmological perturbations has been worked out by Gleyzes \textit{et al.} 
\cite{GLPVprl,GLPV}, based on (a) a Lagrangian depending on the lapse
function and some geometrical scalar quantities emerging in the
Arnowitt-Deser-Misner (ADM) decomposition on the flat Friedmann-Lema\^{\i}%
tre-Robertson-Walker (FLRW) background and (b) the unitary gauge, allowing
to absorb the scalar field perturbation by an adequate time coordinate
choice (the lapse is then associated with the corresponding constant
scalar-field hypersurfaces). The linear perturbation equations contain time
derivatives at second order, although spatial derivatives could be of higher
order (in the Horndeski subclass the latter are also of second order). A
generalization for two scalar fields representing dark matter and dark
energy has been advanced in Ref. \cite{GT}. Another generalization has been
discussed in Ref. \cite{KGT}, referring to perturbations of a spherically
symmetric and static background, treated similarly.

These theories should obey the requirements of

\begin{itemize}
\item[(A)] stability, guaranteed by the avoidance of both scalar ghosts (no
negative kinetic term in the second order Lagrangian governing the evolution
of linear perturbations) and Laplacian instabilities (no negative sound
speed squared),

\item[(B)] agreement with Solar System tests, notably the Vainshtein
mechanism suppressing the propagation of the fifth force inside the Solar
System (no $L_{5}$ contribution to the Horndeski Lagrangian) \cite%
{Vainshtein1,Vainshtein2,Vainshtein3},

\item[(C)] agreement with weak lensing observations (constraints from
deviations from the Newtonian law and light bending by simultaneous fitting
of x-ray and lensing profiles of galaxy clusters) \cite{GalaxyClusters}.
\end{itemize}

The recent detections of gravitational waves from 10 coalescing binary black
holes and one neutron star merger by the LIGO Scientific Collaboration and
Virgo Collaboration \cite{GW1,GW2,GW3,GW4,GW5,GW6,GW7-11} have added new
constraints. On the one hand, the mass of the graviton has been severely
constrained by testing a massive dispersion relation \cite{GRtest}. Then a
wide family of dispersion relations \cite{dispersion} were tested, disruling 
\cite{GW3} Lorentz-violation, Ho\v{r}ava-Lifsic theories, certain extra
dimensional, multifractal theories, doubly special relativity and setting an
even harder constraint on the graviton mass at $5.0\times 10^{-23}$ eV/c$^{2}
$\cite{GW12}. On the other hand the small difference in the arrival time of
the gravitational waves from a neutron star coalescence \cite{GW6} and
accompanying $\gamma -$radiation confirmed that the tensorial gravitational
modes propagate with the speed of light within $-3\times 10^{-15}$ and $%
+7\times 10^{-16}$ accuracy \cite{Multimessenger}. By exploring previously
existing analyses on the Laplacian stability and ghost avoidance in
Horndeski theories \cite{stab1,stab2}, from these constraints the $L_{5}$
contribution has been disruled once again, together with the kinetic term
dependence of $L_{4}$ \cite{GWc1,GWc2,GWc3}. A slightly less restrictive
condition emerged for the beyond Horndeski models. Further, three of the
five parameters appearing in the effective theory of dark energy were
severely constrained by combining the gravity wave results with galaxy
cluster observations \cite{GWc4}.

The stability of spherically symmetric, static spacetimes has been discussed
for both the odd \cite{HornOdd} and even modes \cite{HornEven} of the
perturbations in Horndeski theories, also for the odd modes in the beyond
Horndeski theories \cite{KGT}. The latter relied on a double foliation of
spacetime along orthogonal spatial and temporal leaves, developed in Refs. 
\cite{s+1+1a,s+1+1b}. Three independent background dynamical equations were
identified and the conditions for avoidance of ghosts and Laplacian
instabilities of the odd mode perturbations established.

The formalism of the orthogonal double foliation relies on the extensive use
of adapted metric variables, which bear the role of canonical coordinates
and on embedding variables (extrinsic curvatures, normal fundamental forms
and normal fundamental scalars of the 2-surfaces generated by the
intersection of the foliations), some of them emerging as canonical momenta,
others as pure spatial derivatives of the coordinates. The odd sector of
perturbations of spherically symmetric, static spacetimes has been analyzed
in terms of these quantities \cite{KGT}.

Spacetime perturbations can also be discussed through other decomposition
techniques, including: (I) the first order system of 70 coupled differential
equations for 50 independent variables of the black hole perturbation
formalism \`{a} la Chandrasekhar \cite{Chandra}, based on the Newman-Penrose
formalism (an 1+1+1+1 decomposition); (II) the formalism based on the
numerous variables arising from a 2+1+1 decomposition based on kinematical
quantities (optical scalars), supplemented by the electric and magnetic
projection of the Weyl tensor \cite{Kinem1,Kinem2}; (III) a (2+1)+1
decomposition based on the introduction of the quotient space defined by the
orbits of a rotational Killing vector \cite{Sasaki1,Sasaki2}; (IV) a
temporal foliation followed by a further 2+1 slicing to deal with
axisymmetric and stationary configurations \cite{GourgoulhonBonazzola},
generalized later on for a 2+1 foliation of a hypersurface with arbitrary
causal character \cite{Gourgoulhon,GourgoulhonJaramillo}, a technique also
employed in Ref. \cite{Racz} for identifying a hyperbolic system in the
constraint structure, rewritten in terms of the 2+1 decomposition of the
extrinsic curvature of the hypersurfaces explored previously in the
orthogonal double foliation formalism of Ref. \cite{s+1+1a}; (V) the
standard metric perturbation formalism, explored in a spherically symmetric,
static setup in Refs. \ \cite{HornOdd,HornEven}. The advantage of the
orthogonal double foliation formalism over the first two consist in its
substantially reduced number of variables. A comparison with the third and
fourth has been presented in \cite{s+1+1a}. The third relies heavily on the
use of a Killing vector, which is not a necessity for the orthogonal double
foliation. Although the fourth approach contains the same number of metric
variables (9), it does not employ all geometric quantities playing an
essential role in Refs. \cite{s+1+1a,s+1+1b}. In particular, Ref. \cite%
{Gourgoulhon} introduces a second fundamental form combining a set of
dynamical and nondynamical variables explored in Refs. \cite{s+1+1a,s+1+1b},
a normal fundamental form but no normal fundamental scalar. Finally, the
advantage over the metric perturbation formalism is the canonical
(geometrodynamical) interpretation of the variables.\footnote{%
Other spacetime decomposition techniques are also known. Applying the
formalism developed in the seminal monograph \cite{Schouten}, a 2+2 breakup
of the field equations was advanced in Ref. \cite{INFERNOStachel78} with the
aim of identifying the gravitational d.o.f. in the so-called conformal
two-structure (the latter representing the information on how the family of
selected 2-surfaces is embedded in a 3-surface). For the discussion of the
initial value problem Ref. \cite{INFERNO} developed the 2+2 decomposition of
spacetime in detail, based on space-like 2-surfaces $\{S\}$ rigged by a dyad
basis given by their two mutually orthogonal normals (and the respective
orthogonal 3-foliations). Then the covariant derivatives of these normals
were decomposed in terms of the extrinsic curvatures of $\{S\}$, the induced
connection of the timelike 2-surface $\{T\}$ spanned by the dyad basis and
the curvature tensor of $\{T\}$. The Einstein equations were decomposed
accordingly.}

The simplicity of the orthogonal double foliation of Refs. \cite%
{s+1+1a,s+1+1b} however required to waste one gauge d.o.f. for imposing the
orthogonality requirement after the perturbation. This hampered the
discussion of the even modes, carrying an arbitrary function of time, hence
losing their physical interpretation \cite{KGT}.

It is the purpose of the present paper to lift the condition of
orthogonality of the two foliations in order to recover the full power of
gauge fixing and open the way for the discussion of the even mode
perturbations in generic scalar-tensor theories on a spherically symmetric,
static background, complementing the similar discussion of the odd sector.

The paper is organized as follows. In Sec.~\ref{formalism} we develop the
new 2+1+1 decomposition of the spacetime $\mathcal{B}$ based on two
nonorthogonal foliations, one of them temporal ($\mathcal{S}_{t}$,
characterized by constant $t$), the other one spatial ($\mathfrak{M}_{\chi }$%
, with constant $\chi $). This generalizes the formalism of the orthogonal
double foliation of spacetime, developed in Refs.~\cite{s+1+1a,s+1+1b}, by
allowing for a $10$th metric function $\mathcal{N}$. We adapt suitable bases
to both foliations, then give the evolutions along the $\partial /\partial t$
and $\partial /\partial \chi $ congruences (tangent to $\mathfrak{M}_{\chi }$
and $\mathcal{S}_{t}$, respectively) in both bases. Two of the basis vectors
(tangent to the intersection $\Sigma _{t\chi }$ of $\mathcal{S}_{t}$ and $%
\mathfrak{M}_{\chi }$) are common in both bases, while the other two pairs
are related by a Lorentz rotation with angle $\phi =\tanh ^{-1}\left( 
\mathcal{N}/N\right) $, where $N$\ is the lapse function. Another geometric
interpretation of the $10$th metric function arises as the vorticity of the
basis vectors orthogonal to both the hypersurface normals (of the same
basis) and to $\Sigma _{t\chi }$. This is shown here through the discussion
of the algebras of each basis vectors and in the discussion of the
vorticities in the two Appendices.

In Sec. \ref{KabKaKLabLaL} we characterize the embedding in terms of
extrinsic curvatures, normal fundamental forms and normal fundamental
scalars of the hypersurface normals, also introduce the 2+1 decomposed form
of the curvature of their congruences (their nongravitational
accelerations). For the basis vectors orthogonal to them we introduce
similar quantities. We establish the interconnections among all those
geometric quantities. In Sec. \ref{kinematics} we also derive their
connection with the time- and $\chi $-derivatives of the metric functions.
This enables us to select those geometric variables, which bear a dynamical
role, e.g., connected to canonical momenta.

As a first application, we present the Hamiltonian formalism of general
relativity in the 2+1+1 decomposed form in Sec. \ref{Hamiltonian}. We derive
the canonical momenta, then the Hamiltonian and diffeomorphism constraints
and the boundary terms of the action, all in terms of canonical data defined
on $\Sigma _{t\chi }$. We recover previous results of Ref. \cite{s+1+1b}
applying for the orthogonal double foliation in the vanishing $\mathcal{N}$
limit. The nonorthogonality of the foliations also generates new terms.

Then, in Sec.~\ref{Gaugefix} we explore the diffeomorphism gauge freedom for
fixing the perturbations on the static and spherically symmetric background
of beyond Horndeski theories in an unambiguous way. This result opens up the
possibility for the discussion of the even sector of the perturbations.
Although the unambiguous gauge fixing is different from the one employed for
the odd sector in Ref. \cite{KGT}, the results of the stability analysis
presented there are unaffected, as the two sectors decouple.

In Sec.~\ref{conclusec} we present our conclusions. Two Appendices are
devoted to discuss the consequences of the hypersurface orthogonality of the
normal basis vectors and the interpretation of the vorticities of the
complementary basis vectors in terms of the geometric quantities introduced
in the main body of the paper.

We use the abstract index notation throughout the paper. Latin and greek
indices, respectively, denote 4-dimensional spacetime and 3-dimensional
spatial abstract indices. Boldface lower- and uppercase indices
differentiate among 2-dimensional and 4-dimensional basis vectors,
respectively. 4-dimensional quantities will carry a distinguishing tilde
sign, while 3-dimensional quantities a overhat (or reversed overhat) sign.
Tensors defined both on the full spacetime and on lower-dimensional
(hyper)surfaces carry Latin indices, the latter obeying the required
projection conditions. Quantities defined on the background in a
perturbational setup carry an overbar. Round or square brackets on indices
denote symmetrization or antisymmetrization, respectively.

\section{The nonorthogonal 2+1+1 decomposition of spacetime\label{formalism}}

Let $\mathcal{B}$ be a 4-dimensional manifold with metric $\tilde{g}_{ab}$
of Lorentzian signature. We assume the manifold admits both a timelike and a
spacelike foliation. In this section we generalize the formalism of \cite%
{s+1+1a, s+1+1b} by dropping the orthogonality requirement of the foliations 
$\mathcal{S}_{t}$ (with constant time coordinate $t$) and $\mathfrak{M}%
_{\chi }$ (with constant space coordinate $\chi $).

On the tangent space of the doubly-foliable spacetime $\mathcal{B}$ we
introduce the bases $e_{\mathbf{A}}=\left\{ \partial /\partial t,\partial
/\partial \chi ,E_{\mathbf{i}}\right\} $ (with $E_{\mathbf{i}}$ some basis
elements of the tangent space of $\Sigma _{t\chi }$) and its dual $e^{%
\mathbf{B}}=\left\{ dt,d\chi ,E^{\mathbf{j}}\right\} $ on the respective
cotangent space.

Let $n^{a}$ be the (timelike) unit normal to $\mathcal{S}_{t}$, while $m^{a}$
the (spacelike) unit normal to both $n^{a}$ and $\Sigma _{t\chi }$. With
them we introduce the basis $f_{\mathbf{A}}=\left\{ n,m,F_{\mathbf{i}%
}\right\} $ adapted to $\mathcal{S}_{t}$ (with $F_{\mathbf{i}}$ basis
elements of the tangent space of $\Sigma _{t\chi }$) and its dual $f^{%
\mathbf{B}}=\left\{ \bar{n},\bar{m},F^{\mathbf{j}}\right\} $. The
(spacelike) unit normal to $\mathfrak{M}_{\chi }$ is $l^{a}$, while $k^{a}$
denotes the (timelike) unit normal to both $l^{a}$ and $\Sigma _{t\chi }$.
The basis adapted to $\mathfrak{M}_{\chi }$ is $g_{\mathbf{A}}=\left\{
k,l,G_{\mathbf{i}}\right\} $ (where $G_{\mathbf{i}}$ are basis elements of
the tangent space of $\Sigma _{t\chi }$), with $g^{\mathbf{B}}=\left\{ \bar{k%
},\bar{l},G^{\mathbf{j}}\right\} $ its dual.

For simplicity one can chose coordinate basis vectors $E_{\mathbf{i}}=F_{%
\mathbf{i}}=G_{\mathbf{i}}=\partial /\partial y^{\mathbf{i}}$. From the
causal character of the basis vectors and from the duality relations we get%
\begin{eqnarray*}
\bar{n}_{a} &=&-n_{a}~,\quad \bar{m}_{a}=m_{a}~, \\
\bar{k}_{a} &=&-k_{a}~,\quad \bar{l}_{a}=l_{a}~.
\end{eqnarray*}

\subsection{The induced metric}

The 4-metric $\tilde{g}_{ab}$ can be decomposed in two equivalent ways%
\begin{eqnarray}
\tilde{g}_{ab} &=&-n_{a}n_{b}+m_{a}m_{b}+g_{ab}~,  \label{gtildef} \\
\tilde{g}_{ab} &=&-k_{a}k_{b}+l_{a}l_{b}+g_{ab}~.  \label{gtildeg}
\end{eqnarray}%
As usual $\tilde{g}_{a}^{b}\equiv \delta _{a}^{b}$, while the mixed form of
the induced metric $g_{ab}$ projects to $\Sigma _{t\chi }$. With this
projection, both covariant derivatives and Lie derivatives along a
congruence $V^{a}$ of any 4-dimensional\ tensor $\tilde{T}%
_{b_{1}...b_{r}}^{a_{1}...a_{r}}$ could be projected onto $\Sigma _{t\chi }$%
: 
\begin{equation}
D_{a}\tilde{T}_{b_{1}...b_{q}}^{a_{1}...a_{r}}\equiv
g_{a}^{c}g_{c_{1}}^{a_{1}}...g_{c_{r}}^{a_{r}}g_{b_{1}}^{d_{1}}...g_{b_{q}}^{d_{q}}%
\tilde{\nabla}_{c}\tilde{T}_{d_{1}...d_{q}}^{c_{1}...c_{r}}~,
\label{projkovd}
\end{equation}%
\begin{equation}
\mathfrak{L}_{\mathbf{V}}\tilde{T}_{b_{1}...b_{q}}^{a_{1}...a_{r}}\equiv
g_{c_{1}}^{a_{1}}...g_{c_{r}}^{a_{r}}g_{b_{1}}^{d_{1}}...g_{b_{q}}^{d_{q}}%
\mathfrak{\tilde{L}}_{\mathbf{V}}\tilde{T}_{d_{1}...d_{q}}^{c_{1}...c_{r}}~.
\label{projLied}
\end{equation}%
We note that whenever $\tilde{T}_{b_{1}...b_{q}}^{a_{1}...a_{r}}$ is a
projected object onto $\Sigma _{t\chi }$, the expression $D_{a}\tilde{T}%
_{b_{1}...b_{q}}^{a_{1}...a_{r}}$ is exactly the covariant derivative in $%
\Sigma _{t\chi }$ (which annihilates $g_{ab}$), while $\mathfrak{L}_{\mathbf{%
V}}\tilde{T}_{b_{1}...b_{q}}^{a_{1}...a_{r}}$ describes an evolution along
the congruence $V^{a}$ (it represents the partial derivative with respect to
the adapted coordinate $v$, thus $\mathbf{V}=\partial /\partial v$, where $v$
could be either $t$ or $\chi $). Otherwise they become but notations, as
they fail to obey the Leibniz rule \cite{s+1+1a}).

\subsection{Evolutions in the $f_{\mathbf{A}}$ basis}

\begin{figure}[th]
\includegraphics[height=8cm,angle=0]{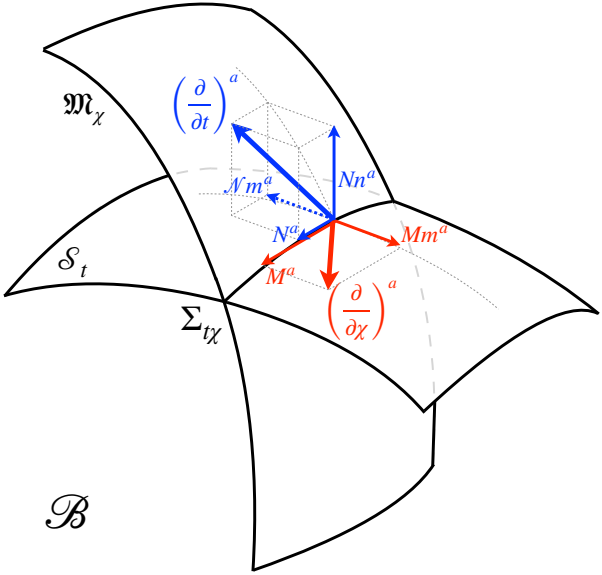} 
\caption{The decomposition of the temporal and radial evolution vectors in
the $f_{\mathbf{A}}$ basis. (For visualization purposes a negative $\mathcal{%
N}$ was chosen.)}
\label{fa}
\end{figure}

The first two elements of the coordinate basis $e_{\mathbf{A}}$,
representing evolution vectors can be generically decomposed in the $f_{%
\mathbf{A}}$ basis as%
\begin{eqnarray}
\left( \frac{\partial }{\partial t}\right) ^{a} &=&Nn^{a}+N^{a}+\mathcal{N}%
m^{a}~,  \label{ddt} \\
\left( \frac{\partial }{\partial \chi }\right) ^{a} &=&Mm^{a}+M^{a}+\mathcal{%
M}n^{a}~.  \label{ddchi}
\end{eqnarray}%
Here $N^{a}$ and $M^{a}$ ($\mathcal{N}$ and $\mathcal{M}$) are the
components of the 3-dimensional shift vectors along (orthogonal to) $\Sigma
_{t\chi }$, while $N$ and $M$ represent lapse type functions of the
respective evolutions. Together with the 3 independent components of $g_{ab}$
there seem to be $11$ gravitational variables at this stage, but their
number will be reduced to $10$. Indeed, the duality relation $\left\langle
dt,\partial /\partial \chi \right\rangle =0$ implies $\mathcal{M}=0$, which
in turn implies through Eq. (\ref{ddchi}) that $\partial /\partial \chi $ is
tangent to $\mathcal{S}_{t}$, see also Fig. \ref{fa}.

From the rest of the duality relations $\left\langle e^{\mathbf{B}},e_{%
\mathbf{A}}\right\rangle =\delta _{\mathbf{A}}^{\mathbf{B}}$ one gets%
\begin{eqnarray}
\bar{n} &=&Ndt~,  \notag \\
\bar{m} &=&\mathcal{N}dt+Md\chi ~,  \notag \\
F^{\mathbf{j}} &=&N^{\mathbf{j}}dt+M^{\mathbf{j}}d\chi +E^{\mathbf{j}}~.
\label{nmF}
\end{eqnarray}

As $\partial /\partial t$ is timelike and $N^{a}$ spacelike, the
inequalities 
\begin{equation*}
N^{2}-\mathcal{N}^{2}>g_{ab}N^{a}N^{b}\geq 0~
\end{equation*}%
hold, while $\partial /\partial t$ lying in the future light cone implies $%
N>0$.

We conclude this subsection by giving in Table \ref{fA} the algebra of the
basis vectors $f_{\mathbf{A}}$. As expected from the Frobenius theorem, the
basis vectors $\left\{ m,F_{\mathbf{i}}\right\} $ span the tangent space of $%
\mathcal{S}_{t}$, while from the dual form of the Frobenius theorem the
fourth basis vector $n^{a}$ turns out vorticity-free (also shown explicitly
in Appendix \ref{hypersurfaceorth}). The same type of reasoning yields that $%
m^{a}$ has vorticity (as the component along $m^{a}$ of the $\left[ n,F_{%
\mathbf{j}}\right] $ bracket is nonvanishing, hence the vectors $\left\{
n,F_{\mathbf{i}}\right\} $ do not span a hypersurface). This vorticity is
given in Appendix \ref{vortykm} and disappears together with $\mathcal{N}$
in the orthogonal foliation limit employed in Refs. \cite{s+1+1a, s+1+1b}.
Hence the vorticity of the basis vector $m^{a}$ is generated by the
nonorthogonality of the two foliations. 
\begin{table*}[tbph]
\begin{center}
\begin{tabular}{cccc}
& \multicolumn{1}{||c}{$\left[ n,m\right] ^{a}$} & \multicolumn{1}{|c}{$%
\left[ n,F_{\mathbf{j}}\right] ^{a}$} & \multicolumn{1}{|c}{$\left[ m,F_{%
\mathbf{j}}\right] ^{a}$} \\ \hline\hline
$n^{a}$ & \multicolumn{1}{||c}{$\frac{1}{M}\left[ \partial _{\chi }\left(
\ln N\right) -\frac{1}{M}M^{\mathbf{j}}\partial _{\mathbf{j}}\left( \ln
N\right) \right] $} & \multicolumn{1}{|c}{$\partial _{\mathbf{j}}\left( \ln
N\right) $} & \multicolumn{1}{|c}{$0$} \\ 
$m^{a}$ & \multicolumn{1}{||c}{$\frac{1}{MN}\left[ -\partial _{t}M+\partial
_{\chi }\mathcal{N}+N^{\mathbf{j}}\partial _{\mathbf{j}}M-M^{\mathbf{j}%
}\partial _{\mathbf{j}}\mathcal{N}\right] $} & \multicolumn{1}{|c}{$\frac{M}{%
N}\partial _{\mathbf{j}}\left( \frac{\mathcal{N}}{M}\right) $} & 
\multicolumn{1}{|c}{$\partial _{\mathbf{j}}\left( \ln M\right) $} \\ 
$F_{\mathbf{i}}^{a}$ & \multicolumn{1}{||c}{$\frac{1}{MN}\left( -\partial
_{t}M^{\mathbf{i}}+\partial _{\chi }N^{\mathbf{i}}+N^{\mathbf{j}}\partial _{%
\mathbf{j}}M^{\mathbf{i}}-M^{\mathbf{j}}\partial _{\mathbf{j}}N^{\mathbf{i}%
}\right) $} & \multicolumn{1}{|c}{$\frac{1}{N}\left[ \partial _{\mathbf{j}%
}N^{\mathbf{i}}-\frac{\mathcal{N}}{M}\partial _{\mathbf{j}}M^{\mathbf{i}}%
\right] $} & \multicolumn{1}{|c}{$\frac{\partial _{\mathbf{j}}M^{\mathbf{i}}%
}{M}$} \\ 
&  &  & 
\end{tabular}%
\end{center}
\caption{The algebra of the basis vectors $f_{\mathbf{A}}$. The components
of the brackets enlisted in the first line along the vectors in the first
column are given.}
\label{fA}
\end{table*}

\subsection{The role of the $10$th metric variable}

The new element in the formalism as compared with that of Refs. \cite%
{s+1+1a, s+1+1b} is the shift component $\mathcal{N}$, which reestablishes
the number of gravitational variables as $10$, equivalent to the 4-metric
variables.

Straightforward calculations employing also the rest of the duality
relations $\left\langle f^{\mathbf{B}},f_{\mathbf{A}}\right\rangle =\delta _{%
\mathbf{A}}^{\mathbf{B}}=\left\langle g^{\mathbf{B}},g_{\mathbf{A}%
}\right\rangle $ and Eqs. (\ref{nmF}) lead to the relation between the two
adapted bases 
\begin{equation*}
\begin{pmatrix}
\bar{k} \\ 
\bar{l}%
\end{pmatrix}%
=%
\begin{pmatrix}
\mathfrak{c} & -\mathfrak{s} \\ 
-\mathfrak{s} & \mathfrak{c}%
\end{pmatrix}%
\begin{pmatrix}
\bar{n} \\ 
\bar{m}%
\end{pmatrix}%
~,
\end{equation*}%
(where $\mathfrak{s}=\sinh \phi $, $\mathfrak{c}=\cosh \phi $) and 
\begin{equation}
\begin{pmatrix}
k^{a} \\ 
l^{a}%
\end{pmatrix}%
=%
\begin{pmatrix}
\mathfrak{c} & \mathfrak{s} \\ 
\mathfrak{s} & \mathfrak{c}%
\end{pmatrix}%
\begin{pmatrix}
n^{a} \\ 
m^{a}%
\end{pmatrix}%
~,  \label{klnm}
\end{equation}%
thus in the form of a Lorentz-rotation. Its angle is defined by 
\begin{equation}
\mathcal{N}=N\tanh \mathfrak{\phi }~.  \label{Phi}
\end{equation}%
This represents the second geometric interpretation of the $10$th metric
variable (beyond the vorticity of $m$).

\subsection{Evolutions in the $g_{\mathbf{A}}$ basis}

With the Lorentz rotations given in the previous subsection it is easy to
express the evolution vectors in the $g_{\mathbf{A}}$ basis:%
\begin{eqnarray}
\left( \frac{\partial }{\partial t}\right) ^{a} &=&\frac{N}{\mathfrak{c}}%
k^{a}+N^{a}~,  \label{ddtg} \\
\left( \frac{\partial }{\partial \chi }\right) ^{a} &=&M\left( -\mathfrak{s}%
k^{a}+\mathfrak{c}l^{a}\right) +M^{a}~.  \label{ddchig}
\end{eqnarray}%
Remarkably, the evolution vector $\partial /\partial t$ has no component
along $l$, hence it is tangent to $\mathfrak{M}_{\chi }$, see also Fig. \ref%
{ga}.

\begin{figure}[th]
\includegraphics[height=8cm,angle=0]{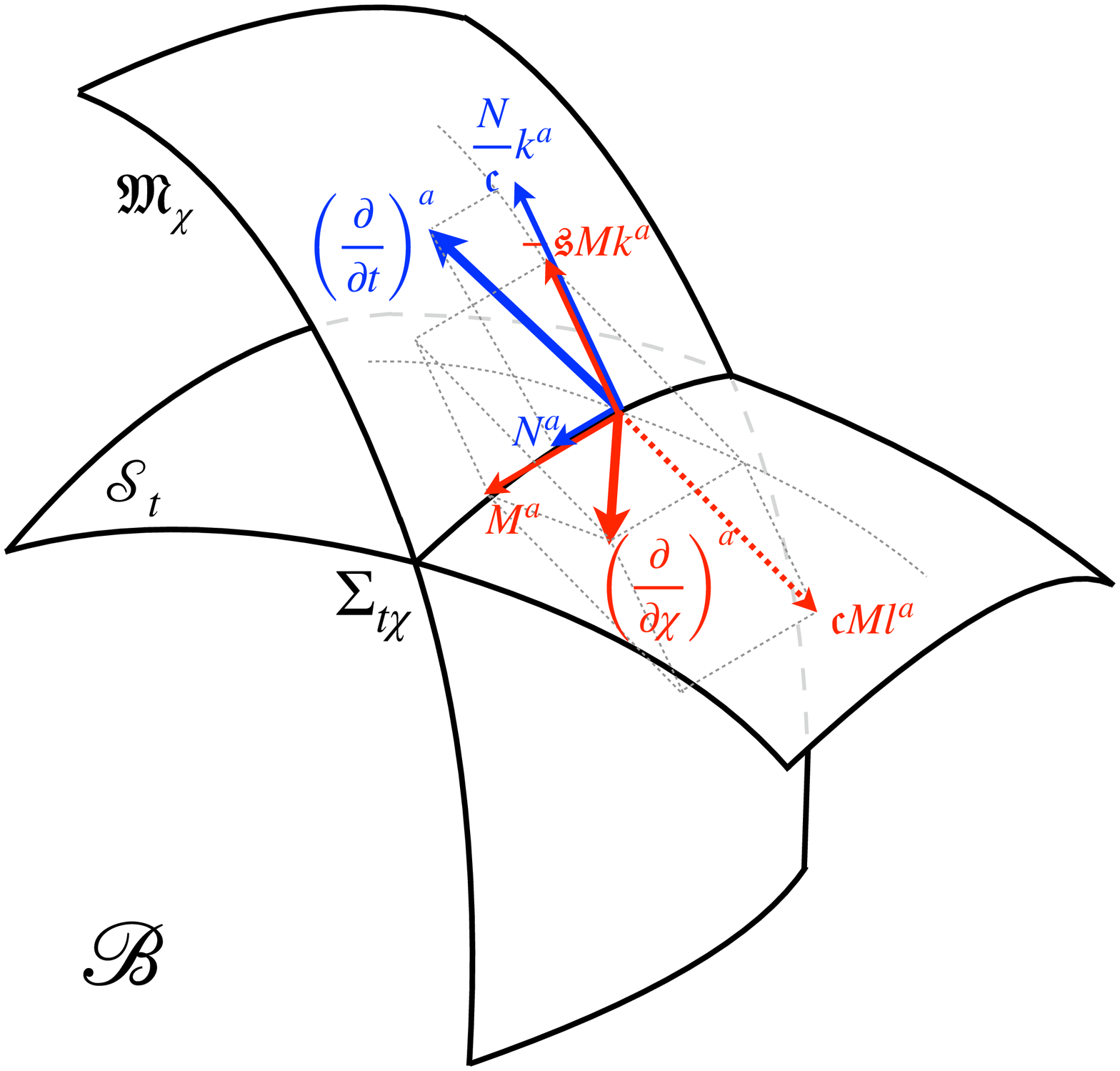} 
\caption{The decomposition of the temporal and radial evolution vectors in
the $g_{\mathbf{A}}$ basis. (For visualization purposes a negative $%
\mathfrak{s}$ was chosen.)}
\label{ga}
\end{figure}

Further exploring the duality relations one gets%
\begin{eqnarray}
\bar{k} &=&-M\mathfrak{s}d\chi +\frac{N}{\mathfrak{c}}dt~,  \notag \\
\bar{l} &=&M\mathfrak{c}d\chi ~,  \notag \\
G^{\mathbf{j}} &=&N^{\mathbf{j}}dt+M^{\mathbf{j}}d\chi +E^{\mathbf{j}}=F^{%
\mathbf{j}}~.  \label{Gkoordb}
\end{eqnarray}%
The algebra of the basis vectors $g_{\mathbf{A}}$ is presented in Table \ref%
{gA}. Again, from the Frobenius theorem, the basis vectors $\left\{ k,G_{%
\mathbf{i}}\right\} $ span the tangent space of $\mathfrak{M}_{\chi }$ and
from its dual form the fourth basis vector $l^{a}$ turns out vorticityfree.
The vector $k^{a}$ however has vorticity (as the component along $k^{a}$ of
the $\left[ l,G_{\mathbf{j}}\right] $ bracket is nonvanishing). This
vorticity is given in Appendix \ref{vortykm} and again disappears with $%
\mathcal{N}$ in the orthogonal foliation limit employed in Refs. \cite%
{s+1+1a, s+1+1b}. Hence the vorticity of the basis vector $k^{a}$ is also
generated by the nonorthogonality of the two foliations. Finally we note
that in the orthogonal foliation limit $\mathcal{N}\rightarrow 0$ the
algebras given in Tables \ref{fA}-\ref{gA} coincide and the vorticities of
the basis vectors disappear. 
\begin{table*}[tbph]
\begin{center}
\begin{tabular}{c||c|c|c}
& $\left[ k,l\right] ^{a}$ & $\left[ k,G_{\mathbf{j}}\right] ^{a}$ & $\left[
l,G_{\mathbf{j}}\right] ^{a}$ \\ \hline\hline
$k^{a}$ & $\left\{ \partial _{t}\left( \frac{\mathfrak{s}}{N}\right) -N^{%
\mathbf{j}}\partial _{\mathbf{j}}\left( \frac{\mathfrak{s}}{N}\right) +\frac{%
\mathfrak{s}}{N}\left[ \partial _{t}\ln \left( MN\right) -N^{\mathbf{j}%
}\partial _{\mathbf{j}}\ln \left( MN\right) \right] \right. $ & $\partial _{%
\mathbf{j}}\left( \ln \frac{N}{\mathfrak{c}}\right) $ & $-\frac{N}{\mathfrak{%
c}^{2}M}\partial _{\mathbf{j}}\left( \frac{\mathfrak{sc}M}{N}\right) $ \\ 
& $\left. +\frac{1}{\mathfrak{c} M}\left[ \partial _{\chi }\ln \left( \frac{N%
}{\mathfrak{c}}\right) -M^{\mathbf{j}}\partial _{\mathbf{j}}\ln \left( \frac{%
N}{\mathfrak{c} }\right) \right] \right\} $ &  &  \\ 
$l^{a}$ & $\frac{1}{MN}\left[ -\partial _{t}\left( {{{\mathfrak{c}}}M}%
\right) +N^{\mathbf{j}}\partial _{\mathbf{j}}\left({{\mathfrak{c}}M}\right) %
\right] $ & $0$ & $\partial _{\mathbf{j}}\ln \left( \mathfrak{c}M\right) $
\\ 
$G_{\mathbf{i}}^{a}$ & $\frac{1}{MN}\left[ -\partial _{t}M^{\mathbf{i}%
}+\partial _{\chi }N^{\mathbf{i}}-M^{\mathbf{j}}\partial _{\mathbf{j}}N^{%
\mathbf{i}}+N^{\mathbf{j}}\partial _{\mathbf{j}}M^{\mathbf{i}}\right] $ & $%
\frac{\mathfrak{c}}{N}\left( \partial _{\mathbf{j}}N^{\mathbf{i}}\right) $ & 
$\frac{\mathfrak{s}}{N}\partial _{\mathbf{j}}N^{\mathbf{i}}+\frac{1}{%
\mathfrak{c}M}\partial _{\mathbf{j}}M^{\mathbf{i}}$%
\end{tabular}%
\end{center}
\caption{The algebra of the basis vectors $g_{\mathbf{A}}$. The components
of the brackets enlisted in the first line along the vectors in the first
column are given.}
\label{gA}
\end{table*}

\section{Codimension-2 embedding of $\Sigma _{t\protect\chi }$ \label%
{KabKaKLabLaL}}

In this section we introduce a series of geometrical quantities
characterizing the embedding of $\Sigma _{t\chi }$ and we analyze their
relationship with various coordinate derivatives of the metric variables.

We have defined a total of four normals to the surface $\Sigma _{t\chi }$,
two pairs taken from the bases $f_{\mathbf{A}}$ and $g_{\mathbf{A}}$,
respectively. With each of them we define an extrinsic curvature, as
follows: 
\begin{eqnarray}
K_{ab} &\equiv &D_{a}n_{b}=\frac{1}{2}\mathfrak{L}_{\mathbf{n}}g_{ab}~, 
\notag \\
L_{ab} &\equiv &D_{a}l_{b}=\frac{1}{2}\mathfrak{L}_{\mathbf{l}}g_{ab}~, 
\notag \\
K_{ab}^{\ast } &\equiv &D_{a}k_{b}=\frac{1}{2}\mathfrak{L}_{\mathbf{k}%
}g_{ab}~,  \notag \\
L_{ab}^{\ast } &\equiv &D_{a}m_{b}=\frac{1}{2}\mathfrak{L}_{\mathbf{m}%
}g_{ab}~.  \label{extrcurv}
\end{eqnarray}%
All these tensors are symmetric, as shown in the Appendices \ref%
{hypersurfaceorth} and \ref{vortykm}.

With the two normals to the hypersurfaces we define the normal fundamental
forms of $\Sigma _{t\chi }$ as follows:%
\begin{eqnarray}
\mathcal{K}_{a} &\equiv &g_{a}^{c}m^{d}\tilde{\nabla}_{c}n_{d}=g_{a}^{c}m^{d}%
\tilde{\nabla}_{d}n_{c}~,  \notag \\
\mathcal{L}_{a} &\equiv &-g_{a}^{c}k^{d}\tilde{\nabla}%
_{c}l_{d}=-g_{a}^{c}k^{d}\tilde{\nabla}_{d}l_{c}~.  \label{KLform}
\end{eqnarray}%
Their second expressions arise from the hypersurface-orthogonality of the
basis vectors $n^{a}$ and $l^{a}$, as proven in Appendix \ref%
{hypersurfaceorth}. It is easy to prove that they are related as 
\begin{equation}
\mathcal{L}_{a}=\mathcal{K}_{a}+D_{a}\mathfrak{\phi ~}.  \label{LakapcsKa}
\end{equation}

By contrast, for the vectors $k^{a}$ and $m^{a}$ (which have vorticity) the
similarly defined quantities 
\begin{eqnarray}
\mathcal{K}_{a}^{\ast } &\equiv &g_{a}^{d}l^{c}\tilde{\nabla}_{c}k_{d}~, 
\notag \\
\mathcal{L}_{a}^{\ast } &\equiv &-g_{a}^{d}n^{c}\tilde{\nabla}_{c}m_{d}
\label{KLstarform}
\end{eqnarray}%
do not share this interchangeability property. Similarly, one can prove%
\begin{equation}
\mathcal{L}_{a}^{\ast }=\mathcal{K}_{a}^{\ast }+D_{a}\mathfrak{\phi ~.}
\label{LastarkapcsKastar}
\end{equation}

The differences $\mathcal{K}_{a}^{\ast }-\mathcal{L}_{a}$ and $\mathcal{L}%
_{a}^{\ast }-\mathcal{K}_{a}$ give the nonvanishing components of the
vorticities of $k^{a}$ and $m^{a}$, respectively, as demonstrated in
Appendix \ref{vortykm}.

For the hypersurface-orthogonal vectors $n^{a}$ and $l^{a}$ normal
fundamental scalars 
\begin{eqnarray}
\mathcal{K} &\equiv &m^{d}m^{c}\tilde{\nabla}_{c}n_{d}~,  \notag \\
\mathcal{L} &\equiv &k^{d}k^{c}\tilde{\nabla}_{c}l_{d}~  \label{KLscalar}
\end{eqnarray}%
can be defined. The corresponding quantities for the basis vectors $k^{a}$
and $m^{a}$ are%
\begin{eqnarray}
\mathcal{K}^{\ast } &\equiv &l^{d}l^{c}\tilde{\nabla}_{c}k_{d}~,  \notag \\
\mathcal{L}^{\ast } &\equiv &n^{c}n^{d}\tilde{\nabla}_{c}m_{d}~.
\label{KLstarscalar}
\end{eqnarray}

Finally the two timelike vector congruences have the curvatures
(nongravitational 3-dimensional accelerations):%
\begin{equation}
\hat{\alpha}_{a}\equiv n^{b}\tilde{\nabla}_{b}n_{a}=\mathfrak{a}_{a}-m_{a}%
\mathcal{L}^{\ast }~,  \label{ngyors}
\end{equation}%
\begin{equation}
\hat{\alpha}_{a}^{\ast }\equiv k^{b}\tilde{\nabla}_{b}k_{a}=\mathfrak{a}%
_{a}^{\ast }-l_{a}\mathcal{L}~,  \label{kgyors}
\end{equation}%
the second set of expressions representing their 2+1 decomposed form with
the 2-dimensional acceleration components:%
\begin{eqnarray}
~\mathfrak{a}_{a} &\equiv &g_{a}^{c}n^{b}\tilde{\nabla}_{b}n_{c}~,  \notag \\
\mathfrak{a}_{a}^{\ast } &\equiv &g_{a}^{c}k^{b}\tilde{\nabla}_{b}k_{c}~.
\label{as}
\end{eqnarray}%
Similarly, the spacelike congruences $l^{a}$ and $m^{a}$ have the
3-dimensional curvatures:%
\begin{equation}
\check{\beta}_{a}\equiv l^{b}\tilde{\nabla}_{b}l_{a}=\mathfrak{b}_{a}+k_{a}%
\mathcal{K}^{\ast }~,  \label{lgyors}
\end{equation}%
\begin{equation}
\check{\beta}_{a}^{\ast }\equiv m^{b}\tilde{\nabla}_{b}m_{a}=\mathfrak{b}%
_{a}^{\ast }+n_{a}\mathcal{K~},  \label{mgyors}
\end{equation}%
with the 2-dimensional \textquotedblleft acceleration\textquotedblright\
components: 
\begin{eqnarray}
~~~~\mathfrak{b}_{a} &\equiv &g_{a}^{d}l^{c}\tilde{\nabla}_{c}l_{d}~,  \notag
\\
\mathfrak{b}_{a}^{\ast } &\equiv &g_{a}^{d}m^{c}\tilde{\nabla}_{c}m_{d}~.
\label{bs}
\end{eqnarray}

With the above-introduced quantities the 2+1+1 decomposition of the
covariant derivatives of the normals to $\Sigma _{t\chi }$ in the bases they
belong is%
\begin{eqnarray}
\tilde{\nabla}_{a}n_{b} &=&K_{ab}+2m_{(a}\mathcal{K}_{b)}+m_{a}m_{b}\mathcal{%
K}+n_{a}m_{b}\mathcal{L}^{\ast }  \notag \\
&&-n_{a}\mathfrak{a}_{b}~,  \label{nfelb} \\
\tilde{\nabla}_{a}l_{b} &=&L_{ab}+2k_{(a}\mathcal{L}_{b)}+k_{a}k_{b}\mathcal{%
L}+l_{a}k_{b}\mathcal{K}^{\ast }  \notag \\
&&+l_{a}\mathfrak{b}_{b}~,  \label{lfelb} \\
\tilde{\nabla}_{a}k_{b} &=&K_{ab}^{\ast }+l_{a}\mathcal{K}_{b}^{\ast }+l_{b}%
\mathcal{L}_{a}+l_{a}l_{b}\mathcal{K}^{\ast }+k_{a}l_{b}\mathcal{L}  \notag
\\
&&-k_{a}\mathfrak{a}_{b}^{\ast }~,  \label{kfelb} \\
\tilde{\nabla}_{a}m_{b} &=&L_{ab}^{\ast }+n_{a}\mathcal{L}_{b}^{\ast }+n_{b}%
\mathcal{K}_{a}+n_{a}n_{b}\mathcal{L}^{\ast }+m_{a}n_{b}\mathcal{K}  \notag
\\
&&+m_{a}\mathfrak{b}_{b}^{\ast }~.  \label{mfelb}
\end{eqnarray}%
For deriving Eqs. (\ref{nfelb}), (\ref{lfelb}) we have also employed the
second equalities (\ref{KLform}). The structure of Eqs. (\ref{kfelb}), (\ref%
{mfelb}) is slightly different due to the vorticities of the vectors $k^{a}$
and $l^{a}$.

The geometric quantities defined in this subsection are not all independent.
This should be obvious as the two bases are related by a Lorentz-rotation.
By tedious but straightforward algebra we expressed all starry quantities in
terms of unstarred ones and $\mathfrak{\phi }$ (or $\mathcal{N}$). For
example the extrinsic curvatures defined with the basis vectors of the two
bases are related by a rotation matrix with angle $\psi =\arccos \left(
1/\cosh \phi \right) $ as:%
\begin{equation}
\left( 
\begin{array}{c}
K_{ab}^{\ast } \\ 
L_{ab}^{\ast }%
\end{array}%
\right) =\left( 
\begin{array}{cc}
1/\mathfrak{c} & \mathfrak{s}/\mathfrak{c} \\ 
-\mathfrak{s}/\mathfrak{c} & 1/\mathfrak{c}%
\end{array}%
\right) \left( 
\begin{array}{c}
K_{ab} \\ 
L_{ab}%
\end{array}%
\right) ~.  \label{KstarLstarKL}
\end{equation}

\begin{figure}[th]
\includegraphics[height=8cm,angle=0]{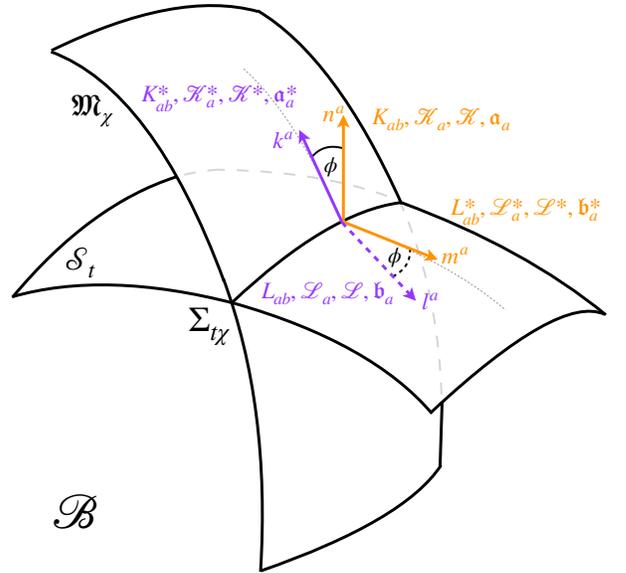} 
\caption{The geometric embedding variables.}
\label{bas}
\end{figure}

The geometric quantities characterizing the embedding are summarized on Fig. %
\ref{bas} while the full set of interdependencies are given them in Table %
\ref{kapcsolatok}.

\begin{table*}[tbph]
\begin{center}
\begin{tabular}{|l|l|}
\hline
$K_{ab}^{\ast }=\frac{1}{\mathfrak{c}}\left( K_{ab}+\mathfrak{s}%
L_{ab}\right) $ & $L_{ab}^{\ast }=\frac{1}{\mathfrak{c}}\left( L_{ab}-%
\mathfrak{s}K_{ab}\right) $ \\ \hline
$\mathcal{K}_{a}^{\ast }=\mathcal{K}_{a}+\frac{\mathfrak{s}}{\mathfrak{c}}%
\left( \mathfrak{a}_{a}+\mathfrak{b}_{a}\right) $ & $\mathcal{L}_{a}^{\ast }=%
\mathcal{L}_{a}+\frac{\mathfrak{s}}{\mathfrak{c}}\left( \mathfrak{a}_{a}+%
\mathfrak{b}_{a}\right) $ \\ \hline
$\mathcal{K}^{\ast }=\frac{1}{\mathfrak{c}}\left( \mathcal{K}-\mathfrak{s}%
\mathcal{L}\right) +\frac{1}{\mathfrak{c}^{2}}\left( l^{a}-\mathfrak{s}%
n^{a}\right) \tilde{\nabla}_{a}\mathfrak{\phi }$ & $\mathcal{L}^{\ast }=%
\frac{1}{\mathfrak{c}}\left( \mathfrak{s}\mathcal{K}+\mathcal{L}\right) +%
\frac{1}{\mathfrak{c}^{2}}\left( \mathfrak{s}l^{a}+n^{a}\right) \tilde{\nabla%
}_{a}\mathfrak{\phi }$ \\ \hline
$\mathfrak{a}_{a}^{\ast }=\mathfrak{a}_{a}+\frac{\mathfrak{s}}{\mathfrak{c}}%
\left( \mathcal{K}_{a}-\mathcal{L}_{a}\right) =\mathfrak{a}_{a}-\frac{%
\mathfrak{s}}{\mathfrak{c}}D_{a}\mathfrak{\phi }$ & $\mathfrak{b}_{a}^{\ast
}=\mathfrak{b}_{a}+\frac{\mathfrak{s}}{\mathfrak{c}}\left( \mathcal{L}_{a}-%
\mathcal{K}_{a}\right) =\mathfrak{b}_{a}+\frac{\mathfrak{s}}{\mathfrak{c}}%
D_{a}\mathfrak{\phi }$ \\ \hline
\end{tabular}%
\end{center}
\caption{The relations among starred and unstarred geometric quantities
characterizing the embedding of $\Sigma _{t\protect\chi }$.}
\label{kapcsolatok}
\end{table*}

Note that the notations were introduced such that in the particular case $%
\mathcal{N}=0$ all starry quantities transform into the corresponding
unstarred ones (e.g., $K_{ab}^{\ast }$ becomes $K_{ab}$). Further, as in
that case the vorticities of the basis vectors $k^{a}$ and $m^{a}$ vanish, $%
\mathcal{L}_{a}=\mathcal{K}_{a}$ (as explored in Refs. \cite{s+1+1a, s+1+1b}%
) follows.

\section{Kinematics and geometric embedding\label{kinematics}}

In this section we establish the relations of the temporal and spatial
derivatives of the metric variables $\left\{ g_{ab},~M^{a},~M\right\} $ to
the geometric quantities $\left\{ K^{ab},~\mathcal{K}_{a},~\mathcal{K}%
\right\} $, $\left\{ L^{ab},~\mathcal{L}_{a},~\mathcal{L}\right\} $, $%
\left\{ K^{\ast ab},~\mathcal{K}_{a}^{\ast },~\mathcal{K}^{\ast }\right\} $
and $\left\{ L^{\ast ab},~\mathcal{L}_{a}^{\ast },~\mathcal{L}^{\ast
}\right\} $ characterizing the embedding. These will be used later in the
derivation of the Hamiltonian formulation of GR from the Einstein-Hilbert
action.

Bearing in mind that both the coordinate derivatives along time and $\chi $
and the extrinsic curvatures are projected Lie derivatives, we find for the
extrinsic curvature in the two bases

\begin{eqnarray}
K_{ab} &=&\frac{1}{N}\!\left[ \frac{1}{2}\partial _{t}g_{ab}\!-\!D_{(a}N_{b)}%
\right] \!-\!\frac{\mathfrak{s}}{M\mathfrak{c}}\!\left[ \frac{1}{2}\partial
_{\chi }g_{ab}\!-\!D_{(a}M_{b)}\right] ,  \notag \\
L_{ab}^{\ast } &=&\frac{1}{M}\left[ \frac{1}{2}\partial _{\chi
}g_{ab}-D_{(a}M_{b)}\right] \,  \label{KabLabstar}
\end{eqnarray}%
and 
\begin{eqnarray}
L_{ab} &=&\frac{\mathfrak{s}}{N}\!\left[ \frac{1}{2}\partial
_{t}g_{ab}\!-\!D_{(a}N_{b)}\right] \!+\!\frac{1}{M\mathfrak{c}}\!\left[ 
\frac{1}{2}\partial _{\chi }g_{ab}\!-\!D_{(a}M_{b)}\right] ,  \notag \\
K_{ab}^{\ast } &=&\frac{\mathfrak{c}}{N}\left[ \frac{1}{2}\partial
_{t}g_{ab}-D_{(a}N_{b)}\right] ~,  \label{KabstarLab}
\end{eqnarray}%
respectively. Only $L_{ab}^{\ast }$ is free from time derivatives of the
induced metric, hence nondynamical.

In order to establish the relation of the rest of the geometric variables
with time and $\chi $-derivatives of the metric variables we employ the
following identity holding for all vectors $V_{\mathbf{I}}$ for which $%
\tilde{g}\left( V_{\mathbf{I}},V_{\mathbf{J}}\right) =$constant: 
\begin{eqnarray}
\tilde{g}(V_{\mathbf{A}},\tilde{\nabla}_{V_{\mathbf{B}}}V_{\mathbf{C}}) &=&%
\tilde{g}\left( \left[ V_{\mathbf{A}},V_{\mathbf{B}}\right] ,V_{\mathbf{C}%
}\right)  \notag \\
&&-\tilde{g}(V_{\mathbf{C}},\tilde{\nabla}_{V_{\mathbf{A}}}V_{\mathbf{B}}).
\label{aux1}
\end{eqnarray}

First we apply this identity for the case $V_{\mathbf{B}}=V_{\mathbf{C}}$,
such that the last term vanishes. Then for the basis vectors $f_{\mathbf{A}}$
and $g_{\mathbf{A}}$ perpendicular to $\Sigma _{t\chi }$ the left-hand sides
are the accelerations $\hat{\alpha}_{a}=\tilde{g}\left( f_{\mathbf{A}},%
\tilde{\nabla}_{\mathbf{n}}\mathbf{n}\right) f_{a}^{\mathbf{A}}$, $\hat{%
\alpha}_{a}^{\ast }=\tilde{g}\left( g_{\mathbf{A}},\tilde{\nabla}_{\mathbf{k}%
}\mathbf{k}\right) g_{a}^{\mathbf{A}}$, $\check{\beta}_{a}=\tilde{g}\left(
g_{\mathbf{A}},\tilde{\nabla}_{\mathbf{l}}\mathbf{l}\right) g_{a}^{\mathbf{A}%
}$ and $\check{\beta}_{a}^{\ast }=\tilde{g}\left( f_{\mathbf{A}},\tilde{%
\nabla}_{\mathbf{m}}\mathbf{m}\right) f_{a}^{\mathbf{A}}$. Calculating the
right-hand sides by exploring the specific components of the Lie brackets
given in Tables \ref{fA} and \ref{gA} and comparing the resulting
expressions with the decompositions given in Eqs. (\ref{ngyors}), (\ref%
{kgyors}), (\ref{lgyors}) and (\ref{mgyors}) we obtain the 2-dimensional
accelerations as projected covariant derivatives%
\begin{eqnarray}
\mathfrak{a}_{a} &=&D_{a}\left( \ln N\right) ~,  \notag \\
\mathfrak{b}_{a}^{\ast } &=&-D_{a}\left( \ln M\right) ~,  \notag \\
\mathfrak{b}_{a} &=&-D_{a}\ln \left( \mathfrak{c}M\right) ~,  \notag \\
\mathfrak{a}_{a}^{\ast } &=&D_{a}\left( \ln \frac{N}{\mathfrak{c}}\right) ~,
\label{accelerations}
\end{eqnarray}%
while the normal fundamental scalars emerge as 
\begin{eqnarray}
\mathcal{K} &=&\frac{1}{MN}\left[ \partial _{t}M-\partial _{\chi }\mathcal{N}%
-N^{a}D_{a}M+M^{a}D_{a}\mathcal{N}\right] ~,  \notag \\
\mathcal{L}^{\ast } &=&-\frac{1}{M}\left[ \partial _{\chi }\left( \ln
N\right) -M^{a}D_{a}\left( \ln N\right) \right] ~,  \notag \\
\mathcal{L} &=&-\mathcal{S}-\frac{1}{\mathfrak{c}M}\left[ \partial _{\chi
}\ln \left( \frac{N}{\mathfrak{c}}\right) -M^{a}D_{a}\ln \left( \frac{N}{%
\mathfrak{c}}\right) \right] ~,  \notag \\
\mathcal{K}^{\ast } &=&\frac{1}{MN}\left[ \partial _{t}\left( \mathfrak{c}%
M\right) -N^{a}D_{a}\left( \mathfrak{c}M\right) \right] ~,
\label{normfundscalars}
\end{eqnarray}%
with%
\begin{eqnarray}
\mathcal{S} &=&\partial _{t}\left( \frac{\mathfrak{s}}{N}\right)
-N^{a}D_{a}\left( \frac{\mathfrak{s}}{N}\right)  \notag \\
&&+\frac{\mathfrak{s}}{N}\left[ \partial _{t}\ln \left( MN\right)
-N^{a}D_{a}\ln \left( MN\right) \right]  \label{calS}
\end{eqnarray}%
(an expression which vanishes for orthogonal foliations).

Next we apply the identity (\ref{aux1}) for $n^{b}\tilde{\nabla}_{b}m_{a}=%
\tilde{g}\left( f_{\mathbf{A}},\tilde{\nabla}_{\mathbf{n}}\mathbf{m}\right)
f_{a}^{\mathbf{A}}$ and $l^{b}\tilde{\nabla}_{b}k_{a}=\tilde{g}\left( g_{%
\mathbf{A}},\tilde{\nabla}_{\mathbf{l}}\mathbf{k}\right) g_{a}^{\mathbf{A}}$%
, respectively, obtaining for the $\Sigma _{t\chi }$ projections 
\begin{eqnarray}
\mathcal{L}_{a}^{\ast } &=&\mathcal{K}_{a}+\frac{M}{N}D_{a}\left( \frac{%
\mathcal{N}}{M}\right) ~,  \label{calLastar} \\
\mathcal{K}_{a}^{\ast } &=&\mathcal{L}_{a}-\frac{N}{\mathfrak{c}^{2}M}%
D_{a}\left( \frac{\mathfrak{sc}M}{N}\right) ~.  \label{calKastar}
\end{eqnarray}%
These can be also derived from the expressions given in Table \ref%
{kapcsolatok} together with Eqs. (\ref{LakapcsKa}) and (\ref{accelerations}%
). Now we have everything at hand to derive the relation of the normal
fundamental forms and metric derivatives. For this we rewrite 
\begin{eqnarray*}
\mathcal{K}_{a} &=&-g_{ab}\left[ m,n\right] ^{b}-\mathcal{L}_{a}^{\ast }~, \\
\mathcal{L}_{a} &=&-g_{ab}\left[ k,l\right] ^{b}-\mathcal{K}_{a}^{\ast }~,
\end{eqnarray*}%
employ the algebras of the basis vectors $f_{\mathbf{A}}$ and $g_{\mathbf{A}%
} $ given in Tables \ref{fA} and \ref{gA}, respectively, together with Eqs. (%
\ref{calLastar}) and (\ref{calKastar}), to obtain%
\begin{eqnarray}
\mathcal{K}^{a} &=&\frac{1}{2MN}\left( \partial _{t}M^{a}-\partial _{\chi
}N^{a}-N^{b}D_{b}M^{a}+M^{b}D_{b}N^{a}\right)  \notag \\
&&-\frac{M}{2N}D^{a}\left( \frac{\mathcal{N}}{M}\right) ~,  \notag \\
\mathcal{L}^{a} &=&\frac{1}{2MN}\left( \partial _{t}M^{a}-\partial _{\chi
}N^{a}-N^{b}D_{b}M^{a}+M^{b}D_{b}N^{a}\right)  \notag \\
&&+\frac{N}{2\mathfrak{c}^{2}M}D^{a}\left( \frac{\mathfrak{sc}M}{N}\right) ~,
\notag \\
\mathcal{K}^{\ast a} &=&\frac{1}{2MN}\left( \partial _{t}M^{a}-\partial
_{\chi }N^{a}-N^{b}D_{b}M^{a}+M^{b}D_{b}N^{a}\right)  \notag \\
&&-\frac{N}{2\mathfrak{c}^{2}M}D^{a}\left( \frac{\mathfrak{sc}M}{N}\right) ~,
\notag \\
\mathcal{L}^{\ast a} &=&\frac{1}{2MN}\left( \partial _{t}M^{a}-\partial
_{\chi }N^{a}-N^{b}D_{b}M^{a}+M^{b}D_{b}N^{a}\right)  \notag \\
&&+\frac{M}{2N}D^{a}\left( \frac{\mathcal{N}}{M}\right) ~.
\label{normfunsvectors}
\end{eqnarray}%
Note that the metric derivatives are related to the normal fundamental
vectors, rather then forms.

From the results of this and of the previous section we can conclude that
the independent metric variables with dynamical role are $\left\{
g_{ab},~M^{a},~M\right\} $ while the embedding variables $\left\{ K^{ab},~%
\mathcal{K}_{a},~\mathcal{K}\right\} $ carry information about their
temporal evolution. The extrinsic curvature $L_{ab}^{\ast }$ being the only
one, which contains no time derivatives, it plays a nondynamical role. Hence
we chose the variables emerging in the $f_{\mathbf{A}}$ basis as
independent, $\left\{ K^{ab},~\mathcal{K}_{a},~\mathcal{K}\right\} $
representing momenta, while$\left\{ L^{\ast ab},~\mathcal{L}^{\ast }\right\} 
$ merely spatial derivatives. All other embedding variables can be expressed
in terms of this set.

\section{Hamiltonian formalism in General Relativity\label{Hamiltonian}}

In this section we present the 2+1+1 decomposed Hamiltonian formalism in
general relativity. As discussed earlier, we employ the $f_{\mathbf{A}}$
basis in the decomposition.

\subsection{The 2+1+1 decomposition of the Einstein-Hilbert action}

We define the 2-dimensional Riemann tensor $R_{abcd}$ of the metric induced
in $\Sigma _{t\chi }$ as 
\begin{equation}
R_{abcd}V^{b}=\left( D_{c}D_{d}-D_{d}D_{c}\right) V_{a}~,
\end{equation}%
which written in terms of the geometric quantities arising in the 2+1+1
decomposition and of the 4-dimensional Riemann tensor leads to the following
Gauss-type identity:%
\begin{equation}
R_{abcd}=g_{a}^{i}g_{b}^{j}g_{c}^{k}g_{d}^{l}\tilde{R}_{ijkl}+2\left(
L_{a[c}^{\ast }L_{d]b}^{\ast }-K_{a[c}K_{d]b}\right) \ .  \label{Gauss}
\end{equation}%
The extrinsic curvatures are those appearing in the $f_{\mathbf{A}}$ basis.
Twice contracting this leads to 
\begin{equation}
R=g^{ik}g^{jl}\tilde{R}_{ijkl}+\left( L^{\ast }\right)
^{2}-K^{2}-L_{ab}^{\ast }L^{\ast ab}+K_{ab}K^{ab}~.  \label{Gauss2a}
\end{equation}%
The first term on the right-hand side is decomposed as 
\begin{eqnarray}
g^{ik}g^{jl}\tilde{R}_{ijkl} &=&\tilde{R}+2\left(
n^{j}n^{l}-m^{j}m^{l}\right) \tilde{R}_{jl}  \notag \\
&&-2n^{i}m^{j}n^{k}m^{l}\tilde{R}_{ijkl}~,  \label{Gauss2b}
\end{eqnarray}%
where 
\begin{eqnarray}
m^{i}n^{j}n^{k}m^{l}\tilde{R}_{ijkl} &=&\mathcal{K}^{k}\left( 2\mathcal{L}%
_{k}^{\ast }+\mathcal{K}_{k}\right) -\left( \mathcal{L}^{\ast }\right)
^{2}+\left( \mathcal{K}\right) ^{2}  \notag \\
&&+\mathfrak{\tilde{L}}_{\mathbf{m}}\mathcal{L}^{\ast }\!+\!\mathfrak{\tilde{%
L}}_{\mathbf{n}}\mathcal{K\!}-\!\frac{D^{i}ND_{i}M}{NM},  \notag \\
n^{j}n^{l}\tilde{R}_{jl} &=&-K^{lb}K_{bl}-\mathcal{L}^{\ast }L^{\ast }-2%
\mathcal{K}^{b}\mathcal{K}_{b}-\left( \mathcal{K}\right) ^{2}  \notag \\
&&+\left( \mathcal{L}^{\ast }\right) ^{2}-\mathfrak{\tilde{L}}_{\mathbf{n}}K-%
\mathfrak{\tilde{L}}_{\mathbf{n}}\mathcal{K}-\mathfrak{\tilde{L}}_{\mathbf{m}%
}\mathcal{L}^{\ast }  \notag \\
&&+\frac{D_{b}D^{b}N}{N}+\frac{D^{b}ND_{b}M}{NM},  \notag \\
m^{j}m^{l}\tilde{R}_{jl} &=&-L^{\ast lb}L_{bl}^{\ast }+2\mathcal{K}_{l}%
\mathcal{L}^{\ast l}-\left( \mathcal{L}^{\ast }\right) ^{2}+\left( \mathcal{K%
}\right) ^{2}  \notag \\
&&+\mathcal{K}K+\mathfrak{\tilde{L}}_{\mathbf{n}}\mathcal{K}-\mathfrak{%
\tilde{L}}_{\mathbf{m}}L^{\ast }+\mathfrak{\tilde{L}}_{\mathbf{m}}\mathcal{L}%
^{\ast }  \notag \\
&&-\left( \frac{D_{b}D^{b}M}{M}+\frac{D^{b}MD_{b}N}{NM}\right) .
\label{Gauss2c}
\end{eqnarray}%
In order to prove the above expressions we have explored the useful
identities%
\begin{equation}
\tilde{\nabla}_{a}\mathfrak{a}^{a}=\frac{D_{a}D^{a}N}{N}+\frac{D^{a}ND_{a}M}{%
NM}~,  \label{nablaa}
\end{equation}%
\begin{equation}
\tilde{\nabla}_{a}\mathfrak{b}^{\ast a}=-\left( \frac{D_{a}D^{a}M}{M}+\frac{%
D^{a}MD_{a}N}{NM}\right) ~,  \label{nablab}
\end{equation}%
and 
\begin{equation}
\tilde{\nabla}_{a}n^{a}=K+\mathcal{K~},\quad \tilde{\nabla}_{a}m^{a}=L^{\ast
}-\mathcal{L}^{\ast }~.  \label{K+K,L-L}
\end{equation}%
With these the twice contracted Gauss relation becomes\footnote{%
By suitably transforming the Lie derivatives this expression becomes
identical with the one obtained for orthogonal double foliations, Eq. (A1)
of Ref. \cite{s+1+1b}, after correcting the coefficient of $\left(
L_{ab}^{\ast }L^{\ast ab}-K_{ab}K^{ab}\right) $ from $-3$ to $+1$ in the
latter.}%
\begin{eqnarray}
R &=&\tilde{R}-K^{2}-K_{ab}K^{ab}+\left( L^{\ast }\right) ^{2}+L_{ab}^{\ast
}L^{\ast ab}-2\mathcal{K}^{b}\mathcal{K}_{b}  \notag \\
&&-2\mathcal{K}\left( K+\mathcal{K}\right) +2\mathcal{L}^{\ast }\left( 
\mathcal{L}^{\ast }-L^{\ast }\right)  \notag \\
&&-2\mathfrak{\tilde{L}}_{\mathbf{n}}\left( K+\mathcal{K}\right) +2\mathfrak{%
\tilde{L}}_{\mathbf{m}}\left( L^{\ast }-\mathcal{L}^{\ast }\right)  \notag \\
&&+2\left[ \frac{D_{a}D^{a}N}{N}+\frac{D_{a}D^{a}M}{M}+\frac{D^{a}MD_{a}N}{NM%
}\right] ~.  \label{Gauss2}
\end{eqnarray}%
Noting that $\sqrt{-\tilde{g}}=NM\sqrt{g}$ the Einstein-Hilbert action 
\begin{eqnarray}
S_{EH} &=&\int dt\int d\chi \int_{\Sigma _{t\chi }}d^{2}x\mathcal{L}%
^{G}~,\quad  \notag \\
\mathcal{L}^{G} &=&\sqrt{-\tilde{g}}\tilde{R}  \label{SEH}
\end{eqnarray}%
can be 2+1+1 decomposed as follows:%
\begin{eqnarray}
&&\mathcal{L}^{G}\left[ \left\{ g_{ab},\!M^{a},\!M\right\} \!;\!\left\{
K^{ab},\!\mathcal{K}_{a},\!\mathcal{K}\right\} \!;\!\left\{ L^{\ast ab},\!%
\mathcal{L}^{\ast }\right\} \!;\!\left\{ N,\!N^{a},\!\mathcal{N}\right\} %
\right]  \notag \\
&=&NM\sqrt{g}\left\{ R+K_{ab}K^{ab}+K^{2}-\left( L^{\ast }\right)
^{2}-L_{ab}^{\ast }L^{\ast ab}\right.  \notag \\
&&+2\mathcal{K}^{a}\mathcal{K}_{a}+2\mathcal{K}\left( K+\mathcal{K}\right) -2%
\mathcal{L}^{\ast }\left( \mathcal{L}^{\ast }-L^{\ast }\right)  \notag \\
&&+2\mathfrak{\tilde{L}}_{\mathbf{n}}\left( K+\mathcal{K}\right) -2\mathfrak{%
\tilde{L}}_{\mathbf{m}}\left( L^{\ast }-\mathcal{L}^{\ast }\right) -2\left[
N^{-1}D_{a}D^{a}N\right.  \notag \\
&&\left. \left. +M^{-1}D_{a}D^{a}M+\left( NM\right) ^{-1}D^{a}MD_{a}N\right]
\right\} ~.  \label{S2+1+1}
\end{eqnarray}%
This form of the action is ready to be employed in the Legendre
transformation.

\subsection{The Legendre transformation}

The action (\ref{S2+1+1}) has to be further transformed in order to derive
the canonical momenta. By employing%
\begin{eqnarray*}
\mathfrak{\tilde{L}}_{\mathbf{n}}\left( K+\mathcal{K}\right) &=&\tilde{\nabla%
}_{a}\left[ n^{a}\left( K+\mathcal{K}\right) \right] -\left( K+\mathcal{K}%
\right) ^{2}~, \\
\mathfrak{\tilde{L}}_{\mathbf{m}}\left( L^{\ast }-\mathcal{L}^{\ast }\right)
&=&\tilde{\nabla}_{a}\left[ m^{a}\left( L^{\ast }-\mathcal{L}^{\ast }\right) %
\right] -\left( L^{\ast }-\mathcal{L}^{\ast }\right) ^{2}~,
\end{eqnarray*}%
we rewrite it in a form explicitly containing all boundary terms (total
divergences): 
\begin{eqnarray}
\mathcal{L}^{G} &=&NM\sqrt{g}\left\{ R+K_{ab}K^{ab}-K^{2}-2K\mathcal{K}+2%
\mathcal{K}^{a}\mathcal{K}_{a}\right.  \notag \\
&&\left. -L_{ab}^{\ast }L^{\ast ab}+L^{\ast 2}-2\mathcal{L}^{\ast }L^{\ast
}+2\!\left( NM\right) ^{-1}\!\!D^{a}MD_{a}N\right.  \notag \\
&&\left. -2\tilde{\nabla}_{a}\left[ \hat{\alpha}^{a}-\check{\beta}^{\ast
a}-n^{a}K+m^{a}L^{\ast }\right] \right\} ~.
\end{eqnarray}%
This contains expressions of the metric variables $\left\{
g_{ab},M^{a},M\right\} $, geometric quantities $\left\{ K^{ab},\mathcal{K}%
_{a},\mathcal{K}\right\} $ containing their time derivatives, purely spatial
derivatives $\left\{ L^{\ast ab},\mathcal{L}^{\ast }\right\} $ [see Eqs. (%
\ref{KabLabstar}),(\ref{KabstarLab})]; the lapse and shift components $%
\left\{ N,N^{a},\mathcal{N}\right\} $ and total divergences. The latter do
not contribute to the dynamics, hence can be omitted when calculating the
canonical momenta: 
\begin{eqnarray}
\pi ^{ab} &=&\frac{\partial \mathcal{L}^{G}}{\partial \dot{g}^{ab}}=\sqrt{g}M%
\left[ K^{ab}-g^{ab}\left( K+\mathcal{K}\right) \right] ~,  \notag \\
p_{a} &=&\frac{\partial \mathcal{L}^{G}}{\partial \dot{M}^{a}}=2\sqrt{g}%
\mathcal{K}_{a}~,  \notag \\
p &=&\frac{\partial \mathcal{L}^{G}}{\partial \dot{M}}=-2\sqrt{g}K~.
\label{momenta}
\end{eqnarray}%
With them we rewrite the Lagrangian density once again with the aim to
manifestly obtain the Liouville-form. This is achieved by transforming (the
double of) the terms quadratic in the set $\left\{ K^{ab},\mathcal{K}_{a},%
\mathcal{K}\right\} $ in the Lagrangian density into expressions linear in
the time derivatives of $\left\{ g_{ab},M^{a},M\right\} $. After extensive
calculations we obtain%
\begin{eqnarray}
\mathcal{L}^{G} &=&\pi ^{ab}\dot{g}_{ab}+p_{a}\dot{M}^{a}+p\dot{M}-\mathcal{H%
}^{G}  \notag \\
&&+\mathcal{L}_{t}^{G}+\mathcal{L}_{\chi }^{G}+\mathcal{L}_{D}^{G}~,
\label{alreadyHam}
\end{eqnarray}%
where 
\begin{equation}
\mathcal{H}^{G}=N\mathcal{H}_{\perp }^{G}+N^{a}\mathcal{H}_{a}^{G}+\mathcal{%
NH}_{\mathcal{N}}^{G}  \label{Hdecomp}
\end{equation}%
is the vacuum gravitational Hamiltonian density in GR, a linear combination
of the products of the Lagrange multipliers $\left\{ N,N^{a},\mathcal{N}%
\right\} $ with the Hamiltonian constraint\footnote{%
This expression reproduces Eq. (13a) of Ref. \cite{s+1+1b} after correcting
the misprints in the signs of the second and third term.}: 
\begin{eqnarray}
\mathcal{H}_{\perp }^{G} &=&\sqrt{g}\left\{ M\left( -R-3L^{\ast
ab}L_{ab}^{\ast }+L^{\ast 2}+K_{ab}K^{ab}\right. \right.  \notag \\
&&\left. +2\mathcal{K}_{a}\mathcal{K}^{a}-K^{2}-2K\mathcal{K}\right)
+2g^{ab}\partial _{\chi }L_{ab}^{\ast }  \notag \\
&&\left. -2M^{a}D_{a}L^{\ast }-4L_{ab}^{\ast
}D^{a}M^{b}+2D^{a}D_{a}M\right\} ,  \label{HamCK}
\end{eqnarray}%
(\textquotedblleft angular\textquotedblright ) diffeomorphism constraints
along $\Sigma _{t\chi }$: 
\begin{eqnarray}
\mathcal{H}_{a}^{G} &=&-2\sqrt{g}\left\{ D_{b}\left[ K_{~a}^{b}M-Mg_{~a}^{b}%
\left( K+\mathcal{K}\right) \right] +KD_{a}M\right.  \notag \\
&&\left. +\mathcal{K}_{a}ML^{\ast }+\partial _{\chi }\mathcal{K}%
_{a}-\!M^{b}D_{b}\mathcal{K}_{a}-\!\mathcal{K}_{b}D_{a}M^{b}\right\} ,
\label{diffCK}
\end{eqnarray}%
and along $m^{a}$ (\textquotedblleft radial\textquotedblright\
diffeomorphism constraint):%
\begin{eqnarray}
\mathcal{H}_{\mathcal{N}}^{G} &=&-2\sqrt{g}\left[ M\left( L^{\ast }\mathcal{K%
}-L_{ab}^{\ast }K^{ab}\right) +MD_{a}\mathcal{K}^{a}\right.  \notag \\
&&\left. +2\mathcal{K}^{a}D_{a}M-\partial _{\chi }K+M^{a}D_{a}K\right] ,
\label{diffNCK}
\end{eqnarray}%
respectively, finally the terms%
\begin{eqnarray}
\mathcal{L}_{t}^{G} &=&2\partial _{t}\left[ \sqrt{g}M\left( K+\mathcal{K}%
\right) \right] ~,  \notag \\
\mathcal{L}_{\chi }^{G} &=&2\partial _{\chi }\left[ \sqrt{g}\left( N\mathcal{%
L}^{\ast }-N_{a}\mathcal{K}^{a}-\mathcal{NK}\right) \right] ~,  \notag \\
\mathcal{L}_{D}^{G} &=&-2\sqrt{g}D_{a}\left[ MD^{a}N+N^{b}\left(
MK_{~b}^{a}-M^{a}\mathcal{K}_{b}\right) \right.  \notag \\
&&\left. +NM^{a}\mathcal{L}^{\ast }+\mathcal{N}\left( M\mathcal{K}^{a}-M^{a}%
\mathcal{K}\right) \right]  \label{boundaryK}
\end{eqnarray}%
are boundary contributions. Employing the inverses 
\begin{eqnarray}
K^{ab} &=&\frac{1}{M\sqrt{g}}\left( \pi ^{ab}-\frac{\pi }{2}g^{ab}\right) -%
\frac{p}{4\sqrt{g}}g^{ab}~,  \notag \\
\mathcal{K}_{a} &=&\frac{1}{2\sqrt{g}}p_{a}~,  \notag \\
\mathcal{K} &\mathcal{=}&\frac{1}{4\sqrt{g}}\left( p-\frac{2\pi }{M}\right)
~,  \label{KKK}
\end{eqnarray}%
of Eqs. (\ref{momenta}) and introducing Lie derivatives by remembering that
the momenta are tensor densities\footnote{%
For an arbitrary tensor density $\mathcal{F}=f\sqrt{g}$ (where $f$ is a
tensor) its Lie derivative along $M^{a}$ is $\mathfrak{L}_{\mathbf{M}}%
\mathcal{F}=D_{a}\left( \mathcal{F}M^{a}\right) $.}, all expressions can be
rewritten in terms of the set of canonical coordinates $\left\{
g_{ab},M^{a},M\right\} $ and canonical momenta $\left\{ \pi
^{ab},p_{a},p\right\} $ as follows:%
\begin{eqnarray}
\mathcal{H}_{\perp }^{G} &=&\sqrt{g}\left[ -M\left( R+3L^{\ast
ab}L_{ab}^{\ast }-L^{\ast 2}\right) +2D^{a}D_{a}M\right.  \notag \\
&&\left. +2g^{ab}\left( \partial _{\chi }-\mathfrak{L}_{\mathbf{M}}\right)
L_{ab}^{\ast }\right] +\frac{1}{M\sqrt{g}}\left( \pi _{ab}\pi ^{ab}-\frac{%
\pi ^{2}}{2}\right)  \notag \\
&&+\frac{M}{\sqrt{g}}\left( \frac{1}{2}p_{a}p^{a}+\frac{1}{8}p^{2}-\frac{\pi
p}{2M}\right) ~,  \label{HamC} \\
\mathcal{H}_{a}^{G} &=&-2D_{b}\pi _{a}^{b}+pD_{a}M-\left( \partial _{\chi }-%
\mathfrak{L}_{\mathbf{M}}\right) p_{a}~,  \label{diffC} \\
\mathcal{H}_{\mathcal{N}}^{G} &=&2L_{ab}^{\ast }\pi
^{ab}-2p^{a}D_{a}M-MD_{a}p^{a}  \label{diffNC} \\
&&-\left( \partial _{\chi }-\mathfrak{L}_{\mathbf{M}}\right) p~.  \notag
\end{eqnarray}%
The constraints (\ref{HamC}) and (\ref{diffC}) fully agree with the
respective ones of Ref. \cite{s+1+1b}, while the last constraint (\ref%
{diffNC}) is new, emerging only in the nonorthogonal double foliation.

Similarly, the boundary terms emerge as:%
\begin{eqnarray}
\mathcal{L}_{t}^{G} &=&-\partial _{t}\left( \pi +\frac{Mp}{2}\right) ~, 
\notag \\
\mathcal{L}_{\chi }^{G} &=&\partial _{\chi }\left[ 2\sqrt{g}N\mathcal{L}%
^{\ast }-N_{a}p^{a}+\mathcal{N}\left( \frac{\pi }{M}-\frac{p}{2}\right) %
\right] ~,  \notag \\
\mathcal{L}_{D}^{G} &=&-D_{a}\left\{ 2\sqrt{g}\left( MD^{a}N+NM^{a}\mathcal{L%
}^{\ast }\right) \right.  \notag \\
&&+N^{b}\left[ 2\pi _{~b}^{a}-\left( \pi +\frac{Mp}{2}\right)
g_{~b}^{a}-M^{a}p_{b}\right]  \notag \\
&&\left. +\mathcal{N}\left[ Mp^{a}+M^{a}\left( \frac{\pi }{M}-\frac{p}{2}%
\right) \right] \right\} ~.  \label{boundary}
\end{eqnarray}%
Unlike the constraints, the boundary terms on the spatial infinity are
modified by new terms proportional to $\mathcal{N}$.

Following the same steps, the time derivatives of the canonical coordinates
can be expressed from Eqs. (\ref{KabLabstar}), (\ref{normfundscalars}) and (%
\ref{normfunsvectors}) as follows:%
\begin{eqnarray}
\dot{g}_{ab} &=&\frac{N}{M\sqrt{g}}\left[ 2\pi _{ab}-\left( \pi +\frac{Mp}{2}%
\right) g_{ab}\right]  \notag \\
&&+\mathfrak{L}_{\mathbf{N}}g_{ab}+\frac{\mathcal{N}}{M}\left( \partial
_{\chi }-\mathfrak{L}_{\mathbf{M}}\right) g_{ab}~,  \notag \\
\dot{M}^{a} &=&\frac{MN}{\sqrt{g}}p^{a}+\left( \partial _{\chi }-\mathfrak{L}%
_{\mathbf{M}}\right) N^{a}+MD^{a}\mathcal{N}-\mathcal{N}D^{a}M~,  \notag \\
\dot{M} &=&\frac{MN}{4\sqrt{g}}\left( p-\frac{2\pi }{M}\right) +\mathfrak{L}%
_{\mathbf{N}}M+\left( \partial _{\chi }-\mathfrak{L}_{\mathbf{M}}\right) 
\mathcal{N~}.  \label{HamCoord}
\end{eqnarray}%
These are but the evolution equations of the canonical coordinates, thus
half of the canonical equations. Note that all of them contain terms with $%
\mathcal{N}$, the rest of the terms agreeing with those derived for the
orthogonal case in Ref. \cite{s+1+1b}.

\subsection{Canonical equations}

In order to simplify the presentation, we introduce the notations $%
g^{A}\equiv \left\{ g_{ab},M^{a},M\right\} $ for the set of canonical
coordinates, $\pi _{A}\equiv \left\{ \pi ^{ab},p_{a},p\right\} $ for the
canonical momenta, and $y=\left\{ y^{1},y^{2}\right\} $ for the coordinates
adapted to $\Sigma _{t\chi }$. The 2+1+1 decomposed Hamiltonian identified
in the previous subsection is 
\begin{equation}
H^{G}=\int d\chi \int dy\mathcal{H}^{G}\left( \chi ,y\right) ~.
\end{equation}%
Time derivatives of the canonical variables emerge as functional derivatives
of the Hamiltonian:%
\begin{eqnarray}
\dot{g}^{A} &=&\frac{\delta H^{G}}{\delta \pi _{A}\left( \chi ,y\right) }~,
\label{gApont} \\
\dot{\pi}_{A} &=&-\frac{\delta H^{G}}{\delta g^{A}\left( \chi ,y\right) }~.
\label{piApont}
\end{eqnarray}%
It can be verified that Eq. (\ref{gApont}) reproduces the set of equations
of motion (\ref{HamCoord}). Next we calculate Eq. (\ref{piApont}) in detail.
Lengthy but straightforward computations lead to the second set of canonical
equations: 
\begin{eqnarray}
\dot{\pi}^{ab} &=&N\mathcal{S}^{ab}+N\mathcal{V}^{ab}-NM\sqrt{g}\mathcal{L}%
^{\ast }(L^{\ast ab}-L^{\ast }g^{ab})  \notag \\
&&+\sqrt{g}\left[ MD^{a}D^{b}N-g^{ab}MD^{c}D_{c}N\right.  \notag \\
&&\left. -g^{ab}\left( D_{c}N\right) \left( D^{c}M\right) +g^{ab}(\partial
_{\chi }-\mathfrak{L}_{\mathbf{M}})(N\mathcal{L}^{\ast })\right]  \notag \\
&&+\mathfrak{L}_{\mathbf{N}}\pi ^{ab}-\left[ \frac{\mathcal{N}\pi ^{ab}}{%
M^{2}}\left( \partial _{\chi }-\mathfrak{L}_{\mathbf{M}}\right) +\mathcal{N}%
p^{(a}D^{b)}\right] M  \notag \\
&&+\left[ \frac{\pi ^{ab}}{M}\left( \partial _{\chi }-\mathfrak{L}_{\mathbf{M%
}}\right) +Mp^{(a}D^{b)}\right] \mathcal{N}  \notag \\
&&+\frac{\mathcal{N}}{M}\left( \partial _{\chi }-\mathfrak{L}_{\mathbf{M}%
}\right) \pi ^{ab}\mathcal{~},  \label{dotpiab}
\end{eqnarray}%
\begin{eqnarray}
\dot{p}_{a} &=&N\mathcal{V}_{a}-2\sqrt{g}[L_{ba}^{\ast }{}D^{b}N+D_{a}(N%
\mathcal{L}^{\ast })]+\mathfrak{L}_{\mathbf{N}}p_{a}~  \notag \\
&&-\frac{2\mathcal{N}}{M}D^{b}\pi _{ba}+\frac{2\mathcal{N}}{M^{2}}\pi
_{ba}D^{b}M  \notag \\
&&+\left( pg_{ab}-\frac{2}{M}\pi _{ab}\right) D^{b}\mathcal{N}~,
\label{padot}
\end{eqnarray}%
\begin{eqnarray}
\dot{p} &=&N\mathcal{S}+N\mathcal{V}-2\sqrt{g}(L^{\ast }\mathcal{L}^{\ast
}+D_{a}D^{a}N)+\mathfrak{L}_{\mathbf{N}}p  \notag \\
&&+\mathcal{N}\left( \frac{2}{M}\pi ^{ab}L_{ab}^{\ast }-D_{a}p^{a}\right)
-2p^{a}D_{a}\mathcal{N~}.  \label{pdot}
\end{eqnarray}%
Here $\mathcal{S}^{ab}$ and $\mathcal{S}$ are 
\begin{eqnarray}
\mathcal{S}^{ab} &=&-\frac{2}{M\sqrt{g}}\left( \pi ^{a}{}_{c}\pi ^{bc}-\frac{%
\pi }{2}\pi ^{ab}\right)  \notag \\
&&+\frac{1}{2M\sqrt{g}}\left( \pi _{cd}\pi ^{cd}-\frac{\pi ^{2}}{2}\right)
g^{ab}  \notag \\
&&-\frac{M}{4\sqrt{g}}g^{ab}\left( \frac{\pi p}{M}-p_{c}p^{c}-\frac{p^{2}}{4}%
\right)  \notag \\
&&+\frac{1}{2\sqrt{g}}\left( p\pi ^{ab}+Mp^{a}p^{b}\right) ~,
\end{eqnarray}%
\newline
\begin{eqnarray}
\mathcal{S} &=&\frac{1}{\sqrt{g}M^{2}}\left( \pi _{ab}\pi ^{ab}-\frac{\pi
^{2}}{2}\right)  \notag \\
&&-\frac{1}{2\sqrt{g}}\left( p_{a}p^{a}+\frac{p^{2}}{4}\right) ~,  \label{S}
\end{eqnarray}%
while $\mathcal{V}^{ab},~\mathcal{V}_{a},$ and $\mathcal{V}$ represent the
tensorial, vectorial, and scalar projections of the force term of the $%
\left( s+1\right) $-dimensional scalar curvature potential, given in Ref. 
\cite{s+1+1b}: 
\begin{eqnarray}
\mathcal{V}^{ab} &=&-M\sqrt{g}\left( G^{ab}+2L^{\ast ac}L^{\ast
b}{}_{c}-L^{\ast }L^{\ast ab}\right)  \notag \\
&&+\frac{M}{2}\sqrt{g}\left( 3L^{\ast cd}L_{cd}^{\ast }-L^{\ast 2}\right)
g^{ab}  \notag \\
&&+\sqrt{g}\left( g^{ac}g^{bd}-g^{ab}g^{cd}\right) \left( \partial /\partial
\chi -\mathfrak{L}_{\mathbf{N}}\right) L_{cd}^{\ast }  \notag \\
&&+\sqrt{g}(D^{a}D^{b}M-g^{ab}D^{c}D_{c}M)~,
\end{eqnarray}%
\begin{equation}
\mathcal{V}_{a}=-2\sqrt{g}\left( D^{b}L_{ba}^{\ast }-D_{b}L^{\ast }\right) ~,
\end{equation}%
\begin{equation}
\mathcal{V}=\sqrt{g}(R+L_{ab}^{\ast }L^{\ast ab}-L^{\ast 2})~.
\end{equation}%
The canonical equations given by Eqs. (\ref{HamCoord}) and (\ref{dotpiab}-%
\ref{pdot}) are the generalizations of Eqs. (29a-30c) of Ref. \cite{s+1+1b}
for the case of nonorthogonal double foliation of the 4-dimensional
spacetime.

\section{Gauge transformations and fixing in perturbations of spherically
symmetric, static black holes in generic scalar-tensor theories\label%
{Gaugefix}}

In GR the perturbations of the spherically symmetric, static spacetime have
been discussed both for the odd \cite{ReggeWheeler} and for the even parity
sectors \cite{Zerilli}. The 10 metric functions were analyzed by employing a
2+1+1 decomposition based on the temporal and radial direction and a further
decomposition of the metric perturbation into spherical harmonics and its
derivatives. As result of the choice of polar coordinates, 8 metric
perturbations survived, however suitably adapting the remaining
diffeomorphism freedom, the odd sector has been expressed in terms of 2, the
even sector in terms of 4, respectively \cite{ReggeWheeler}.

When we consider generic scalar perturbations of spherically symmetric,
static black holes in those scalar-tensor gravitational theories, which
avoid Ostrogradsky instabilities, an additional scalar variable pops in,
further complicating the gauge choice. We address this problem in this
section.

All background quantities will be denoted by an overbar and the respective
perturbed quantities will be written as $N=\bar{N}+\delta N$, etc. Due to
the high degree of symmetry of the background we could assume both the
temporal and spatial evolutions perpendicular to $\Sigma _{t\chi }$ (hence $%
\bar{N}^{a}=\bar{M}^{a}=0$), the foliations perpendicular ($\mathcal{\bar{N}}%
=0$) and radial unitary gauge ($\bar{\phi}=\bar{\phi}\left( \chi \right) $),
thus the scalar depending only on the radial coordinate $\chi $. Hence the
perturbed metric to first order becomes 
\begin{eqnarray}
ds^{2} &=&-\left( \bar{N}^{2}+2\bar{N}\delta N\right) dt^{2}+2\bar{M}\delta {%
\mathcal{N}}dtd\chi  \notag \\
&&+2\delta N_{a}dtdx^{a}+\left( \bar{g}_{ab}+\delta g_{ab}\right)
dx^{a}dx^{b}  \notag \\
&&+2\delta M_{a}dx^{a}d\chi +\left( \bar{M}^{2}+2\bar{M}\delta M\right)
d\chi ^{2}\,,  \label{ds1}
\end{eqnarray}%
while the scalar field changes as 
\begin{equation}
\phi =\bar{\phi}\left( \chi \right) +\delta \phi ~.
\end{equation}

Helmholtz-like decompositions on spherically symmetric background hold for
both vectors:%
\begin{equation}
V_{a}=\bar{D}_{a}V_{\mathrm{rotfree}}+{E^{b}}_{a}\bar{D}_{b}V_{\mathrm{%
divfree}}\,,  \label{vectordec}
\end{equation}%
where $E_{ab}=\sqrt{\bar{g}}\,\varepsilon _{ab}$ is the 2-dimensional
Levi-Civita tensor (having zero projections outside the surfaces of
transitivity of the SO(3) symmetry), with $\varepsilon _{ab}$ the
2-dimensional alternating symbol (with the sign convention $\varepsilon
_{\theta \varphi }=1$ when polar coordinates are adapted). A similar
decomposition holds for any symmetric tensor on spherically symmetric
background into scalar, rotationfree and divergencefree parts:

\begin{eqnarray}
S_{ab} &=&S_{ba}=\bar{g}_{ab}S_{\mathrm{scalar}}+\bar{D}_{a}\bar{D}_{b}S_{%
\mathrm{rotfree}}  \notag \\
&&+\frac{1}{2}\left( {E^{c}}_{a}\bar{D}_{c}\bar{D}_{b}+{E^{c}}_{b}\bar{D}_{c}%
\bar{D}_{a}\right) S_{\mathrm{divfree}}\,.  \label{tensordec}
\end{eqnarray}%
The scalar and rotationfree parts in the above decompositions compose the
even sector under parity transformations, while the divergencefree parts
form the odd sector. These sectors decouple. We decompose all metric
perturbations as follows 
\begin{subequations}
\label{perturbdec}
\begin{eqnarray}
\delta N_{a} &=&\bar{D}_{a}P+{E^{b}}_{a}\bar{D}_{b}Q\,,  \label{Nadec} \\
\delta M_{a} &=&\bar{D}_{a}V+{E^{b}}_{a}\bar{D}_{b}W\,,  \label{Madec} \\
\delta g_{ab} &=&\bar{g}_{ab}A+\bar{D}_{a}\bar{D}_{b}B  \label{hdec} \\
&&+\frac{1}{2}\left( {\ E^{c}}_{a}\bar{D}_{c}\bar{D}_{b}+{E^{c}}_{b}\bar{D}%
_{c}\bar{D}_{a}\right) C\,.  \notag
\end{eqnarray}%
In consequence the odd sector contains the variables: 
\end{subequations}
\begin{equation}
Q,W,C\text{ ,}
\end{equation}%
while the even sector is composed of the variables: 
\begin{equation}
P,V,A,B,\delta N,\delta \mathcal{N},\delta M,\delta \phi ~.
\end{equation}

Next we proceed to fix the gauge in an unambiguous manner. For this we need
the transformation of the metric and scalar perturbations under
diffeomorphisms. The transformed quantities will carry an overhat and they
arise as%
\begin{equation}
\mathfrak{L}_{\xi }\tilde{g}_{ab}=\delta \tilde{g}_{ab}-\widehat{\delta 
\tilde{g}_{ab}}~\,,\quad \mathfrak{L}_{\xi }\phi =\delta \phi -\widehat{%
\delta \phi }~\,,
\end{equation}%
where the vector $\xi ^{a}$ is also decomposed into even and odd
contributions%
\begin{equation}
\left( \xi ^{t},\xi ^{\chi },\xi ^{a}=\bar{D}^{a}\xi +E^{ba}\bar{D}_{b}\eta
\right) ~\,,\qquad \left( a=\theta ,\varphi \right) \,.
\end{equation}%
The transformed quantities were given in Ref. \cite{KGT} as: ~

\begin{subequations}
\label{gaugetrans}
\begin{eqnarray}
\widehat{\delta N} &=&\delta N-\bar{N}\dot{\xi ^{t}}-\bar{N^{\prime }}\xi
^{\chi }\,, \\
\widehat{\delta \mathcal{N}} &=&\delta \mathcal{N}-\frac{\bar{N}^{2}}{2\bar{M%
}}{\xi ^{t}}^{\prime }+\frac{\bar{M}}{2}\dot{\xi ^{\chi }}\,, \\
\widehat{\delta M} &=&\delta M+\bar{M}^{\prime }{\xi ^{\chi }}+\bar{M}{\xi
^{\chi }}^{\prime }\,, \\
\widehat{P} &=&P-\bar{N}^{2}{\xi ^{t}}+\dot{\xi}\,, \\
\widehat{Q} &=&Q+\dot{\eta}\,, \\
\widehat{V} &=&V+{\bar{M}^{2}}{\xi ^{\chi }}+{\xi }^{\prime }-\frac{2}{{\chi 
}}\xi \,, \\
\widehat{W} &=&W+{\eta }^{\prime }-\frac{2}{\chi }\eta \,, \\
\widehat{A} &=&A+\frac{2}{\chi }\xi ^{\chi }\,, \\
\widehat{B} &=&B+2\xi \,, \\
\widehat{C} &=&C+2\eta \,, \\
\widehat{\delta \phi } &=&\delta \phi -\bar{\phi}^{\prime }\xi ^{\chi }\,.
\end{eqnarray}%
Here an overdot and a comma denote time derivative and $\chi $-derivative,
respectively. It is immediate to consume the radial diffeomorphism d.o.f.
for maintaining the radial unitary gauge after the perturbation kicks in,
hence $\widehat{\delta \phi }=0$ (e.g. we chose the $\chi $ coordinate on
the perturbed spacetime such that constant $\chi $-hypersurfaces and
constant $\widehat{\phi }$-hypersurfaces coincide) and 
\end{subequations}
\begin{equation}
\xi ^{\chi }=\frac{\delta \phi }{\bar{\phi}^{\prime }}~.  \label{g1}
\end{equation}%
Then we consume the angular diffeomorphisms by rendering the transformation
of the 2-metric to a conformal transformation, hence $\widehat{B}=0=\widehat{%
C}$. This is achieved by the choices%
\begin{equation}
\xi =-\frac{B}{2}\,,\qquad \eta =-\frac{C}{2}\,.  \label{g23}
\end{equation}%
The last, temporal diffeomorphism was employed in Ref. \cite{KGT} to
reinforce the perpendicularity of the two foliations, hence $\widehat{\delta 
\mathcal{N}}=0$ and 
\begin{equation}
\xi ^{t}=\int d\chi \frac{2\bar{M}}{\bar{N}^{2}}\left( \delta \mathcal{N}+%
\frac{\bar{M}}{2}\dot{\xi}^{\chi }\right) +F(t,\theta ,\varphi )\,,
\label{g4orth}
\end{equation}%
which introduced there an arbitrary function depending on all variables, but
the radial one. This did not affect the analysis of the odd sector, as the
function $F$ emerged only in the even sector variables $\widehat{\delta N}$
and $\widehat{P}$, hampering the physical interpretation of the
perturbations. By exploring the freedom of the nonorthogonality of the two
foliations however we do not have to chose $\xi ^{t}$ in this inconvenient
way. Indeed, we could fix $\widehat{P}=0$, achieved by%
\begin{equation}
{\xi ^{t}}{=}\frac{P+\dot{\xi}}{\bar{N}^{2}}\,.  \label{g4}
\end{equation}%
Therefore the analysis of the even sector perturbations can be carried out
unambiguously. In summary, with the gauge choice advanced in this section,
the remaining odd sector variables%
\begin{equation}
\widehat{Q},\widehat{W}  \label{oddvar}
\end{equation}%
are identical with the ones employed in Ref. \cite{KGT}, while the dynamics
of the even sector perturbations is described in terms of the variables%
\begin{equation}
\widehat{V},\widehat{A},\widehat{\delta N},\widehat{\delta \mathcal{N}},%
\widehat{\delta M}~.  \label{evenvar}
\end{equation}

Note that another unambiguous gauge fixing for spherically symmetric, static
black hole perturbations in Horndeski theories (based on a decomposition
into spherical harmonics) has been advanced in Ref. \cite{HornEven}, however
that choice prefers to cancel the perturbation of the 2-metric rather than
preserving the radial unitary gauge. We summarize the various available
gauge choices in Table \ref{gauges}. The closest to the Regge-Wheeler gauge
is the one developed here, which also ensures the radial unitary gauge.

\begin{table*}[tbph]
\begin{center}
\begin{tabular}{l||l|l||l|l|l}
& \multicolumn{2}{c||}{odd perturbations} & \multicolumn{3}{c}{even
perturbations} \\ \cline{2-6}
& vanishing & physical & vanishing & physical & nonvanishing, nonphysical \\ 
\hline\hline
RW & $\widehat{C}=0$ & $\widehat{Q},~\widehat{W}$ & $\widehat{B}=\widehat{P}=%
\widehat{V}=0$ & ~$\widehat{\delta N},$~$\widehat{\delta \mathcal{N}},~%
\widehat{\delta M},~\widehat{A}$ &  \\ 
KMS & $\widehat{C}=0$ & $\widehat{Q},~\widehat{W}$ & $\widehat{B}=\widehat{P}%
=\widehat{A}=0$ & $~\widehat{\delta N},~\widehat{\delta \mathcal{N}},~%
\widehat{\delta M},~\widehat{V},~\widehat{\delta \phi }$ &  \\ 
KGT & $\widehat{C}=0$ & $\widehat{Q},~\widehat{W}$ & $\widehat{B}=\widehat{%
\delta \phi }=0$ & $~\widehat{\delta M},~\widehat{A},\,\widehat{V}$ & $%
\widehat{\delta N},~\widehat{\delta \mathcal{N}},~\widehat{P}$ \\ 
GKG & $\widehat{C}=0$ & $\widehat{Q},~\widehat{W}$ & $\widehat{B}=\widehat{P}%
=\widehat{\delta \phi }=0$ & $~\widehat{\delta N},~\widehat{\delta \mathcal{N%
}},~\widehat{\delta M},~\widehat{A},~\widehat{V}$ & 
\end{tabular}%
\end{center}
\caption{Comparison of the various gauge choices from the literature for the
odd and even sector perturbations (all transcribed in the notations of this
paper). In the absence of the scalar field and employing a decomposition
into spherical harmonics and their derivatives, Regge and Wheeler (RW) have
adopted a unanimous gauge choice leaving 2 odd and 4 even sector metric
perturbations \protect\cite{ReggeWheeler}. Their approach has been
generalized for Horndeski theories by Kobayashi, Motohashi, and Suyama
(KMS), resulting in the same 2 variables for the odd sector \protect\cite%
{HornOdd} and 5 for the even sector (these include the scalar field
perturbation) \protect\cite{HornEven}. Only 3 out of 4 metric perturbation
variables correspond to those of the Regge-Wheeler choice. In the orthogonal
double foliation formalism, Kase, Gergely, and Tsujikawa (KGT) have employed
the Regge-Wheeler gauge for the odd sector and additionally the radial
unitary gauge \protect\cite{KGT}. The price to pay for the orthogonality of
the foliations was an arbitrary function of time appearing in 3 of the even
sector metric perturbation variables (nevertheless the even sector was
beyond the scope of that paper). In our paper (GKG) we advance another
unambiguous gauge choice for scalar-tensor gravity, containing the
Regge-Wheeler gauge for the odd sector and all variables corresponding to
the even sector analysis of Regge and Wheeler, with an additional one (the
even sector part of the metric perturbation $\protect\delta M^{a}$). This
resulted from imposing the radial unitary gauge, which adapts the $\protect%
\chi $-coordinate to absorb the scalar field perturbation. }
\label{gauges}
\end{table*}

\section{Concluding Remarks\label{conclusec}}

Supplementing existing spacetime decomposition techniques, in this paper we
have developed the decomposition along nonorthogonal double foliations.
Suppressing the orthogonality requirement of the formalism of Ref. \cite%
{s+1+1a}, applied successfully in Ref. \cite{KGT} for the analysis of the
odd sector perturbations of beyond-Horndeski theories, but making impossible
the similar discussion of the even sector, the latter restriction is lifted.
The development of the 2+1+1 nonorthogonal decomposition formalism followed
closely its orthogonal counterpart. The metric has tensorial, vectorial and
scalar contributions respective to the intersections of the leaves $\Sigma
_{t\chi }$. The 2-metric $g_{ab}$, radial shift $M^{a}$ and radial lapse $M$
are canonical coordinates, supplemented by the 2-projection $N^{a}$, radial
projection $\mathcal{N}$ of the temporal shift, and the temporal lapse $N$.
With two nonorthogonal foliations, there are two different adapted bases. We
gave the temporal and radial evolutions in terms of the bases associated
with both foliations.

The metric variable $\mathcal{N}$, absent in the orthogonal double foliation
formalism, has been found as being related to (i) the Lorentz rotation with
angle $\phi =\tanh ^{-1}\left( \mathcal{N}/N\right) $, among these bases,
and (ii) the vorticity of the basis vectors orthogonal to both the
hypersurface normals (of the same basis) and to $\Sigma _{t\chi }$. In the
orthogonal limit of vanishing $\mathcal{N}$, the bases coincide and all
basis vectors become hypersurface-orthogonal. The $10$th metric variable $%
\mathcal{N}$ has been reintroduced as a nonorthogonality measure of the
formalism.

Then the codimension-2 embedding of $\Sigma _{t\chi }$ has been
characterized in terms of extrinsic curvatures, normal fundamental forms and
normal fundamental scalars defined for its normals for both bases and their
network of interrelations established. The study of the kinematics of the
canonical data indicated that the basis containing the normal to the spatial
hypersurfaces is more advantageous, as the embedding variables contain fewer
time derivatives. Hence we explored the quantities related to this basis in
the remaining part of the paper.

As a first application of the spacetime decomposition along a nonorthogonal
double foliation we derived the general relativistic vacuum Hamiltonian
dynamics in a fashion similar to Ref. \cite{s+1+1b}, first 2+1+1 decomposing
the curvature scalar in this formalism (and correcting the coefficients in
the respective previous result), then giving both the canonical coordinates $%
g^{A}$ and the canonical momenta $\pi _{A}$ in terms of the introduced
geometrical quantities. Their dynamics has been worked out as canonical
equations involving the Hamiltonian and diffeomorphism constraints. The
expressions derived reproduced all terms emerging in the orthogonal double
foliation formalism, supplemented by new terms containing $\mathcal{N}$.

In Appendix C of Ref. \cite{s+1+1b} it has been shown that the further 2+1
decomposition employed in the canonical 3+1 ADM formalism leads to the 2+1+1
Hamiltonian formalism with orthogonal double foliation presented there. By
relaxing the orthogonality of the foliations, with a similar technique, the
Hamiltonian formalism presented here can also be derived. Notably exploring
the 2+1 decompositions of the 3-dimensional shift%
\begin{equation}
\hat{N}^{a}=N^{a}+\mathcal{N}m^{a}~
\end{equation}%
and of the diffeomorphism constraint%
\begin{equation}
\widehat{\mathcal{H}}_{a}^{G}=\mathcal{H}_{a}^{G}+\mathcal{H}_{\mathcal{N}%
}^{G}m_{a}~
\end{equation}%
the vacuum GR gravitational Hamiltonian density (\ref{Hdecomp}) emerges.
Similarly with the 2+1 decomposition of the induced 3-metric (\ref{gkalap})
and of the 3-dimensional canonical momentum%
\begin{equation}
\hat{\pi}^{ab}=\pi ^{ab}+Mp^{(a}m^{b)}+\frac{M}{2}pm^{a}m^{b}
\end{equation}%
inserted in the respective equations of the standard ADM\ approach the 2+1+1
decomposed action (\ref{alreadyHam})-(\ref{boundaryK}) and Hamiltonian
equations of motion (\ref{HamCoord}) and (\ref{dotpiab})-(\ref{pdot}) of
this paper follow. Hence the 2+1+1 decomposition and the variational
principle commute.

As compared to the treatment of Ref. \cite{s+1+1b} a new constraint emerged
due to nonorthogonality, the radial diffeomorphism constraint. With this we
reestablished the full constraint structure of general relativity, adapted
to the nonorthogonal double foliation. In a canonical quantum gravity theory
the diffeomorphism constraints must annihilate the physical states. Singling
out the radial diffeomorphism constraint through the 2+1 decomposition of
the 3-dimensional diffeomorphism constraint may turn useful in
midisuperspace models \cite{KucharCGW,Torre,BojowaldBook}, where integration
over the angular sector is carried out and the relevant diffeomorphism
constraint is exactly $\mathcal{H}_{\mathcal{N}}^{G}$.

In the last section we proceeded with the second application of our newly
developed formalism, the gauge fixing of generic scalar-tensor gravitational
theories. As compared to Ref. \cite{KGT} an unambiguous gauge fixing has
been achieved. This includes restricting the perturbation of the 2-metric to
a conformal transformation, freezing the evolution of the scalar field by
ensuring the radial unitary gauge both before and after the perturbation,
and suppressing the even modes of the 2-dimensional shift perturbation. This
last step was not possible with the previous assumption of orthogonal double
foliation, explored in Ref. \cite{KGT}.

Our work opens up the perspective for discussion of the perturbations of
spherically symmetric, static black holes in the effective field theory
approach of scalar-tensor gravitational theories in the radial unitary
gauge, with the inclusion of both the even and odd sectors.

\section*{Acknowledgements}

We are thankful for Zolt\'{a}n Kov\'{a}cs for help in comparing our results
with his previous work. This work was supported by the Hungarian National
Research Development and Innovation Office (NKFIH) in the form of the Grant
No. 123996 and has been carried out in the framework of COST actions CA15117
(CANTATA) and CA16104 (GWverse), supported by COST (European Cooperation in
Science and Technology). C.G. was supported by the UNKP-18-2 New National
Excellence Program of the Ministry of Human Capacities of Hungary. Z.K. was
supported by the J\'{a}nos Bolyai Research Scholarship of the Hungarian
Academy of Sciences and by the UNKP-18-4 New National Excellence Program of
the Ministry of Human Capacities of Hungary.

\appendix

\section{Consequences of the hypersurface orthogonality of the basis vectors 
$\left\{ n,l\right\} $\label{hypersurfaceorth}}

We denote the spatial\ and Lorentzian 3-metrics induced on the $\mathcal{S}%
_{t}$ and $\mathfrak{M}_{\chi }$ hypersurfaces, respectively as 
\begin{eqnarray}
\hat{g}_{ab} &=&m_{a}m_{b}+g_{ab}~,  \label{gkalap} \\
\check{g}_{ab} &=&-k_{a}k_{b}+g_{ab}~.  \label{g3}
\end{eqnarray}%
The basis vectors $n^{a}$ ($l^{a}$) being orthogonal to the $\mathcal{S}_{t}$
($\mathfrak{M}_{\chi }$) hypersurfaces, the dual form of the Frobenius
theorem guarantees the vanishing of their 3-dimensional vorticities: 
\begin{eqnarray}
\hat{\omega}_{ab}^{(\mathbf{n})} &\equiv &\hat{g}_{[a}^{c}\hat{g}_{b]}^{d}%
\tilde{\nabla}_{c}n_{d}=0~,  \label{curv3n} \\
\check{\omega}_{ab}^{(\mathbf{l})} &\equiv &\check{g}_{[a}^{c}\check{g}%
_{b]}^{d}\tilde{\nabla}_{c}l_{d}=0~.  \label{curv3l}
\end{eqnarray}%
These can be further 2+1 decomposed as 
\begin{eqnarray}
0 &=&~D_{[a}n_{b]}+m_{[a}g_{b]}^{d}m^{c}\left( \tilde{\nabla}_{c}n_{d}-%
\tilde{\nabla}_{d}n_{c}\right) ~,  \label{ideigl1} \\
0 &=&D_{[a}l_{b]}+k^{c}g_{[a}^{d}k_{b]}\left( \tilde{\nabla}_{c}l_{d}-\tilde{%
\nabla}_{d}l_{c}\right) ~,  \label{ideigl2}
\end{eqnarray}%
the contraction of which with $m^{a}$ and $k^{a}$, respectively leading to
the second type of expressions of the normal fundamental forms, given in
Eqs. (\ref{KLform}).

\begin{table*}[tbph]
\begin{center}
\begin{tabular}{|l|l|}
\hline
$\hat{\omega}_{ab}^{(\mathbf{k})}g_{c}^{b}=0$ & $\check{\omega}_{ab}^{(%
\mathbf{m})}g_{c}^{b}=0$ \\ \hline
$\hat{\omega}_{ab}^{(\mathbf{k})}l^{b}=\frac{1}{2}D_{a}\mathfrak{\phi }-%
\frac{\mathfrak{s}}{2\mathfrak{c}}\left( \mathfrak{a}_{a}+\mathfrak{b}%
_{a}\right) $ & $\check{\omega}_{ab}^{(\mathbf{m})}n^{b}=\frac{1}{2}D_{a}%
\mathfrak{\phi }+\frac{\mathfrak{s}}{2\mathfrak{c}}\left( \mathfrak{a}_{a}+%
\mathfrak{b}_{a}\right) $ \\ \hline
\end{tabular}%
\end{center}
\caption{The 3-dimensional vorticity components in terms of $\protect\phi $
(or $\mathcal{N}$).}
\label{vorticities}
\end{table*}

Projecting Eqs. (\ref{ideigl1}) and (\ref{ideigl2}) to $\Sigma _{t\chi }$
confirms the vanishing of the 2-dimensional vorticities: 
\begin{equation}
\omega _{ab}^{(\mathbf{n})}\equiv D_{[a}n_{b]}=0~,~  \label{curv2n}
\end{equation}%
\begin{equation}
\omega _{ab}^{(\mathbf{l})}\equiv D_{[a}l_{b]}=0~,  \label{curv2l}
\end{equation}%
and symmetry of the extrinsic curvatures 
\begin{eqnarray}
K_{ab} &\equiv &D_{a}n_{b}=D_{(a}n_{b)}=K_{ba}~,  \notag \\
L_{ab} &\equiv &D_{a}l_{b}=D_{(a}l_{b)}=L_{ba}~,  \label{KabLab}
\end{eqnarray}%
which could also be directly checked from the first Eq. (\ref{nmF}) and the
second Eq. (\ref{Gkoordb}), giving the normals%
\begin{eqnarray}
n_{a} &=&-N\tilde{\nabla}_{a}t~,  \notag \\
l_{a} &=&M\mathfrak{c}\tilde{\nabla}_{a}\chi ~.  \label{nala}
\end{eqnarray}%
Inserting these into the definitions (\ref{KabLab}) manifestly symmetric
expressions arise.

\section{Consequences of the vorticity of the basis vectors $\left\{
k,m\right\} $ \label{vortykm}}

We introduce 3-dimensional metrics, which are orthogonal to the basis
vectors $k^{a}$ and $m^{a}$, respectively:%
\begin{eqnarray}
\hat{h}_{ab} &=&l_{a}l_{b}+g_{ab}~,  \label{hkalap} \\
\check{h}_{ab} &=&-n_{a}n_{b}+g_{ab}~.  \label{h3}
\end{eqnarray}%
These metrics are defined on 3-manifolds which are not hypersurfaces, but
rather the manifolds formed by the integral curves of the vector fields $%
k^{a}$ and $m^{a}$. The 3-dimensional vorticity tensors of $k^{a}$ and $%
m^{a} $ do not vanish, as they are not hypersurface-orthogonal:%
\begin{eqnarray}
\hat{\omega}_{ab}^{(\mathbf{k})} &\equiv &\hat{h}_{[a}^{c}\hat{h}_{b]}^{d}%
\tilde{\nabla}_{c}k_{d}\neq 0~,  \label{curv3k} \\
\check{\omega}_{ab}^{(\mathbf{m})} &\equiv &\check{h}_{[a}^{c}\check{h}%
_{b]}^{d}\tilde{\nabla}_{c}m_{d}\neq 0~.  \label{curv3m}
\end{eqnarray}%
Their 2+1 decomposition leads to 
\begin{eqnarray}
\hat{\omega}_{ab}^{(\mathbf{k})} &=&~g_{[a}^{c}g_{b]}^{d}\tilde{\nabla}%
_{c}k_{d}+l_{[a}g_{b]}^{d}l^{c}\left( \tilde{\nabla}_{c}k_{d}-\tilde{\nabla}%
_{d}k_{c}\right) ~,  \label{ideigl3} \\
\check{\omega}_{ab}^{(\mathbf{m})} &=&g_{[a}^{c}g_{b]}^{d}\tilde{\nabla}%
_{c}m_{d}+g_{[a}^{d}n_{b]}n^{c}\left( \tilde{\nabla}_{c}m_{d}-\tilde{\nabla}%
_{d}m_{c}\right) ~.  \label{ideigl4}
\end{eqnarray}%
Projecting Eqs. (\ref{ideigl3}) and (\ref{ideigl4}) to the $\Sigma _{t\chi }$
surfaces leads to the 2-dimensional vorticities:%
\begin{equation}
\omega _{ab}^{(\mathbf{k})}\equiv D_{[a}k_{b]}=0~,~  \label{curv2k}
\end{equation}%
\begin{equation}
\omega _{ab}^{(\mathbf{m})}\equiv D_{[a}m_{b]}=0~,~  \label{curv2m}
\end{equation}%
which (as both $k^{a}$ and $m^{a}$ are orthogonal to the surface $\Sigma
_{t\chi }$) vanish due to the dual form of the Frobenius theorem.
Alternatively, these can be proved directly through the relations%
\begin{eqnarray}
k_{a} &=&M\mathfrak{s}\tilde{\nabla}_{a}\chi -\frac{N}{\mathfrak{c}}\tilde{%
\nabla}_{a}t~,  \notag \\
m_{a} &=&N\frac{\mathfrak{s}}{\mathfrak{c}}\tilde{\nabla}_{a}t~+M\tilde{%
\nabla}_{a}\chi ~,  \label{kama}
\end{eqnarray}%
emerging from the second Eq. (\ref{nmF}) and the first Eq. (\ref{Gkoordb}).
With them the symmetry of the extrinsic curvatures $K_{ab}^{\ast }$ and $%
L_{ab}^{\ast }$ can be readily checked.

\begin{figure}[h]
\includegraphics[height=7cm,angle=0]{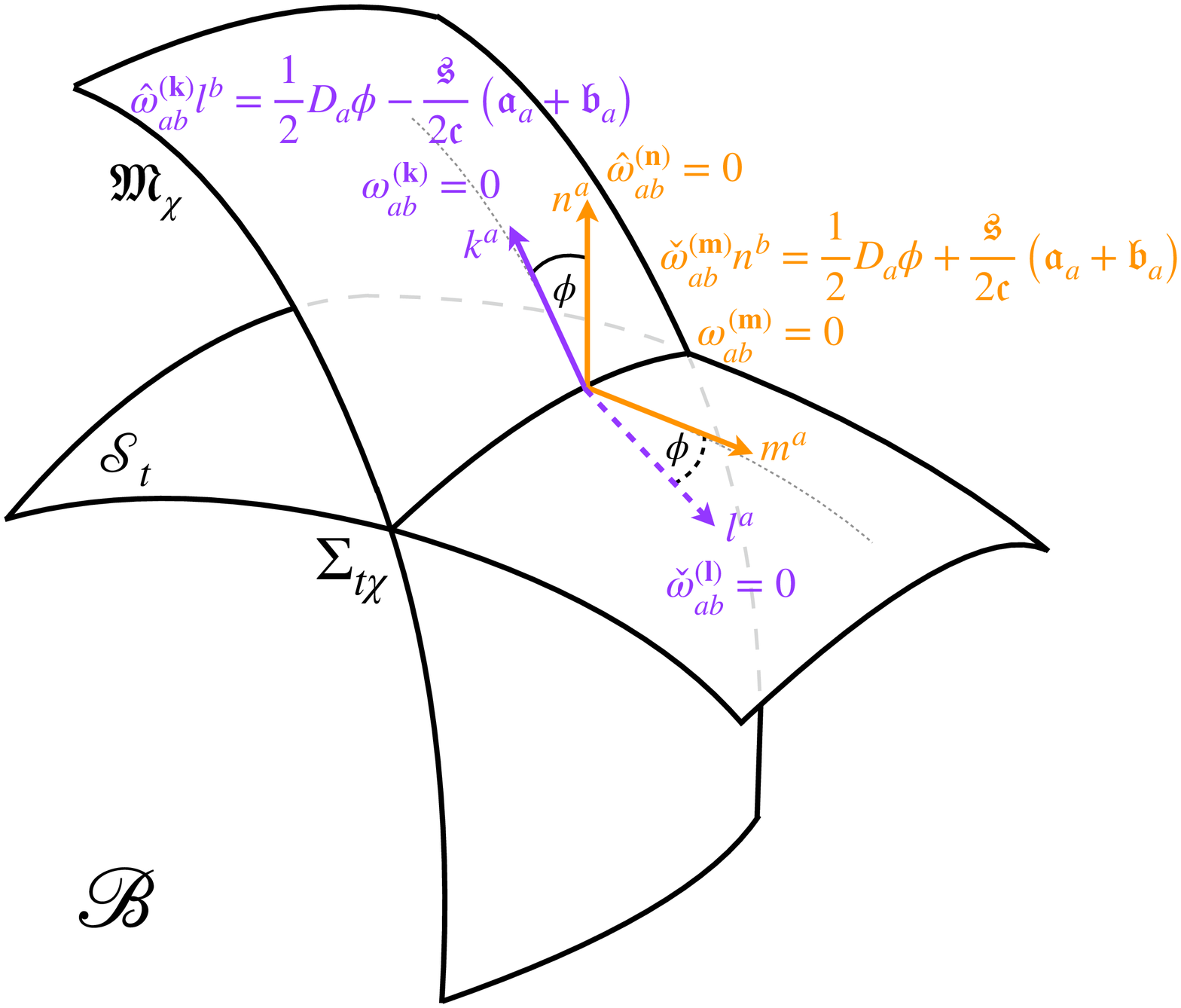} 
\caption{The nonvanishing vorticity components of the basis vectors.}
\label{vort}
\end{figure}

Hence the 3-dimensional vorticities reduce to:%
\begin{eqnarray}
\hat{\omega}_{ab}^{(\mathbf{k})} &=&l_{[a}g_{b]}^{d}l^{c}\left( \tilde{\nabla%
}_{c}k_{d}-\tilde{\nabla}_{d}k_{c}\right) ~,  \notag \\
\check{\omega}_{ab}^{(\mathbf{m})} &=&-n_{[a}g_{b]}^{d}n^{c}\left( \tilde{%
\nabla}_{c}m_{d}-\tilde{\nabla}_{d}m_{c}\right) ~,  \label{om3km}
\end{eqnarray}%
having nonvanishing components only along the normals of the two
hypersurface families. These nonvanishing vorticity components of the basis
vectors are also indicated on Fig. \ref{vort}.

By exploring the definitions (\ref{KLform}) and (\ref{KLstarform}) we get
the expressions of the starry quantities $\mathcal{K}_{a}^{\ast }$ and $%
\mathcal{L}_{a}^{\ast }$ in terms of normal fundamental forms and
nonvanishing 3-dimensional vorticity components:%
\begin{eqnarray}
\mathcal{K}_{a}^{\ast } &=&\mathcal{L}_{a}-2\hat{\omega}_{ab}^{(\mathbf{k}%
)}l^{b}~,  \notag \\
\mathcal{L}_{a}^{\ast } &=&\mathcal{K}_{a}+2\check{\omega}_{ab}^{(\mathbf{m}%
)}n^{b}~.  \label{hosszuvolt}
\end{eqnarray}

By exploring the results given in Table \ref{kapcsolatok}, by
straightforward algebra we can express the 3-dimensional vorticities in
terms of the $10$th metric variable, as given in Table \ref{vorticities}.

\newpage\


\begin{thebibliography}{99}
\bibitem{Horndeski} G.W. Horndeski, Second-order scalar-tensor field
equations in a four-dimensional space, Int. J. Theor. Phys. \textbf{10}, 363
(1974).

\bibitem{Deffayet} C. Deffayet, X. Gao, D. A. Steer, and G. Zahariade, From
k-essence to generalized Galileons, Phys. Rev. D \textbf{84}, 064039 (2011).

\bibitem{GLPVprl} J. Gleyzes, D. Langlois, F. Piazza, and F. Vernizzi,
Healthy theories beyond Horndeski, Phys. Rev. Lett. \textbf{114}, 211101
(2015) [arXiv:1404.6495 [hep-th]].

\bibitem{GLPV} J. Gleyzes, D. Langlois, F. Piazza, and F. Vernizzi,
Essential building blocks of dark energy, J. Cosmol. Astropart. Phys. 08
(2013) 025 [arXiv:1304.4840 [hep-th]].

\bibitem{GT} L. \'{A}. Gergely and S. Tsujikawa, Effective field theory of
modified gravity with two scalar fields: Dark energy and dark matter, Phys.
Rev. D \textbf{89}, 064059 (2014) [arXiv:1402.0553 [hep-th]].

\bibitem{KGT} R. Kase, L. \'{A}. Gergely, and S. Tsujikawa, Effective field
theory of modified gravity on spherically symmetric background: leading
order dynamics and the odd mode perturbations, Phys. Rev. D \textbf{90},
124019 (2014) [arXiv:1406.2402 [hep-th]].

\bibitem{Vainshtein1} R. Kimura, T. Kobayashi, and K. Yamamoto, Vainshtein
screening in a cosmological background in the most general second-order
scalar-tensor theory, Phys. Rev. D \textbf{85}, 024023 (2012)
[arXiv:1111.6749 [astro-ph.CO]].

\bibitem{Vainshtein2} R. Kase and S. Tsujikawa, Screening the fifth force in
the Horndeski's most general scalar-tensor theories,\textit{\ }J. Cosmol.
Astropart. Phys. 08 (2013) 054 [arXiv:1306.6401 [gr-qc]].

\bibitem{Vainshtein3} K. Koyama, G. Niz, and G. Tasinato, Effective theory
for the Vainshtein mechanism from the Horndeski action, Phys. Rev. D \textbf{%
88}, 021502(R) (2013) [arXiv:1305.0279 [hep-th]].

\bibitem{GalaxyClusters} J. Sakstein, H. Wilcox, D. Bacon, K. Koyama, and R.
C. Nichol, Testing Gravity Using Galaxy Clusters: New Constraints on Beyond
Horndeski Theories, J. Cosmol. Astropart. Phys. 07 (2016) 019
[arXiv:1603.06368 [astro-ph.CO]].

\bibitem{GW1} LIGO Scientific and Virgo Collaborations, Observation of
Gravitational Waves from a Binary Black Hole Merger, Phys. Rev. Lett. 
\textbf{116}, 061102 (2016) [arXiv:1602.03837 [gr-qc]].

\bibitem{GW2} LIGO Scientific and Virgo Collaborations, GW151226:
Observation of Gravitational Waves from a 22-Solar-Mass Binary Black Hole
Coalescence, Phys. Rev. Lett. \textbf{116}, 241103 (2016) [arXiv:1606.04855
[gr-qc]].

\bibitem{GW3} LIGO Scientific and Virgo Collaborations, GW170104:
Observation of a 50-Solar-Mass Binary Black Hole Coalescence at Redshift
0.2, Phys. Rev. Lett. \textbf{118}, 221101 (2017) [arXiv:1706.01812 [gr-qc]].

\bibitem{GW4} LIGO Scientific and Virgo Collaborations, GW170608:
Observation of a 19-Solar-Mass Binary Black Hole Coalescence, Astrophys. J.
Lett. \textbf{851}, L35 (2017) [arXiv:1711.05578 [astro-ph.HE]].

\bibitem{GW5} LIGO Scientific and Virgo Collaborations, GW170814: A
Three-Detector Observation of Gravitational Waves from a Binary Black Hole
Coalescence, Phys. Rev. Lett. \textbf{119}, 141101 (2017) [arXiv:1709.09660
[gr-qc]].

\bibitem{GW6} LIGO Scientific and Virgo Collaborations, GW170817:
Observation of Gravitational Waves from a Binary Neutron Star Inspiral,
Phys. Rev. Lett. \textbf{119}, 161101 (2017) [arXiv:1710.05832 [gr-qc]].

\bibitem{GW7-11} LIGO Scientific and Virgo Collaborations, GWTC-1: A
Gravitational-Wave Transient Catalog of Compact Binary Mergers Observed by
LIGO and Virgo during the First and Second Observing Runs, (2018)
[arXiv:181112907 [astro-ph.HE]].

\bibitem{GRtest} LIGO Scientific and Virgo Collaborations, Tests of General
Relativity with GW150914, Phys. Rev. Lett. \textbf{116}, 221101 (2016)
[arXiv:1602.03841 [gr-qc]].

\bibitem{dispersion} S. Mirshekari, N. Yunes, and C. M. Will, Constraining
Lorentz-violating, Modified Dispersion Relations with Gravitational Waves,
Phys. Rev. D \textbf{85}, 024041 (2012) [arXiv:1110.2720 [gr-qc]].

\bibitem{GW12} LIGO Scientific and Virgo Collaborations, Tests of General
Relativity with the Binary Black Hole Signals from the LIGO-Virgo Catalog
GWTC-1, (2019) [arXiv:1602.03837 [gr-qc]].

\bibitem{Multimessenger} LIGO Scientific and Virgo Collaborations, Fermi
Gamma-ray burst monitor, and INTEGRAL, Gravitational Waves and
Gamma-Rays from a Binary Neutron Star Merger: GW170817 and GRB170817A,
Astrophys. J. Lett. \textbf{848}, L13 (2017).

\bibitem{stab1} T. Kobayashi, M. Yamaguchi, and J. Yokoyama, Generalized
G-inflation: Inflation with the most general second-order field equations,
Prog. Theor. Phys. \textbf{126}, 511--529 (2011) [arXiv:1105.5723 [hep-th]].

\bibitem{stab2} A. De Felice and S. Tsujikawa, Conditions for the
cosmological viability of the most general scalar-tensor theories and their
applications to extended Galileon dark energy models, J. Cosmol. Astrophys.
Phys. \textbf{1202}, 007 (2012) [arXiv:1110.3878 [gr-qc]].

\bibitem{GWc1} T. Baker, E. Bellini, P. G. Ferreira, M. Lagos, J. Noller,
and I. Sawicki, Strong constraints on cosmological gravity from GW170817 and
GRB 170817A, Phys. Rev. Lett. \textbf{119}, 251301 (2017) [arXiv:1710.06394
[astro-ph.CO]].

\bibitem{GWc2} J. M. Ezquiaga and M. Zumalac\'{a}rregui, Dark Energy after
GW170817: Dead ends and the road ahead, Phys. Rev. Lett. \textbf{119},
251304 (2017) [arXiv:1710.05901 [astro-ph.CO]].

\bibitem{GWc3} P. Creminelli and F. Vernizzi, Dark Energy after GW170817 and
GRB170817A, Phys. Rev. Lett. \textbf{119}, 251302 (2017) [arXiv:1710.05877
[astro-ph.CO]].

\bibitem{GWc4} J. Sakstein and B. Jain, Implications of the Neutron Star
Merger GW170817 for Cosmological Scalar-Tensor Theories, Phys.Rev.Lett. 
\textbf{119}, 251303 (2017) [arXiv:1710.05893 [astro-ph.CO]].

\bibitem{HornOdd} T. Kobayashi, H. Motohashi and T. Suyama, Black hole
perturbation in the most general scalar-tensor theory with second-order
field equations I: The odd-parity sector, Phys. Rev. D \textbf{85}, 084025
(2012) [arXiv:1202.4893 [gr-qc]].

\bibitem{HornEven} T. Kobayashi, H. Motohashi and T. Suyama, Black hole
perturbation in the most general scalar-tensor theory with second-order
field equations II: the even-parity sector, Phys. Rev. D \textbf{89}, 084042
(2014) [arXiv:1402.6740 [gr-qc]].

\bibitem{s+1+1a} L. \'{A}. Gergely and Z. Kov\'{a}cs, Gravitational dynamics
in s+1+1 dimensions, Phys. Rev. D \textbf{72}, 064015 (2005)
[arXiv:gr-qc/0507020].

\bibitem{s+1+1b} Z. Kov\'{a}cs and L. \'{A}. Gergely, Gravitational dynamics
in s+1+1 dimensions II. Hamiltonian theory, Phys. Rev. D \textbf{77}, 024003
(2008) [arXiv:0709.2131 [gr-qc]].

\bibitem{Chandra} S. Chandrasekhar, \textit{\textquotedblleft The
Mathematical Theory of Black Holes\textquotedblright }, (Oxford Clarendon,
1983).

\bibitem{Kinem1} C. A. Clarkson and R. K. Barrett, Covariant Perturbations
of Schwarzschild Black Holes, Class. Quant. Grav. \textbf{20}, 3855 (2003)
[gr-qc/0209051].

\bibitem{Kinem2} C. Clarkson, A covariant approach for perturbations of
rotationally symmetric spacetimes, Phys. Rev. D \textbf{76}, 104034 (2007)
[arXiv:0708.1398 [gr-qc]].

\bibitem{Sasaki1} K. Maeda, M. Sasaki, T. Nakamura, and S. Miyama, A New
Formalism of the Einstein Equations for Relativistic Rotating Systems, Prog.
Theor. Phys. \textbf{63}, 719 (1980).

\bibitem{Sasaki2} M. Sasaki, [(2+1)+1]-Formalism of General Relativity in 
\textit{Problems of Collapse and Numerical Relativity}, edited by D. Bancel
and M. Signore (Reidel, Dordrecht, 1984), pp. 203-220.

\bibitem{GourgoulhonBonazzola} E. Gourgoulhon and S. Bonazzola, Noncircular
axisymmetric stationary spacetimes, Phys. Rev. D \textbf{48}, 2635 (1993).

\bibitem{Gourgoulhon} E. Gourgoulhon, Generalized Damour-Navier-Stokes
equation applied to trapping horizons, Phys. Rev. D \textbf{72}, 104007
(2005) [arXiv:gr-qc/0508003].

\bibitem{GourgoulhonJaramillo} E. Gourgoulhon and J. L. Jaramillo, A 3+1
perspective on null hypersurfaces and isolated horizons, Phys. Rep. \textbf{%
423}, 159-294 (2006) [arXiv:gr-qc/0503113].

\bibitem{Racz} I. R\'{a}cz, Constraints as evolutionary systems, Class.
Quantum Grav. \textbf{33} 015014 (2016) [arXiv:1508.01810 [gr-qc]].

\bibitem{Schouten} J. A. Schouten, \textit{Ricci-Calculus} (Springer Verlag,
Berlin, 1954).

\bibitem{INFERNOStachel78} R. A. d'Inverno and J. Stachel, Conformal
two-structure as the gravitational degrees of freedom in general relativity,
J. Math. Phys. (N.Y.) \textbf{19}, 2447 (1978).

\bibitem{INFERNO} R. A. d'Inverno and J. Smallwood, Covariant 2+2
formulation of the initial-value problem in general relativity,\textbf{\ }%
Phys. Rev. D \textbf{22}, 1233 (1980).

\bibitem{ReggeWheeler} T. Regge and J. A. Wheeler, Stability of a
Schwarzschild Singularity, Phys. Rev. \textbf{108}, 1063 (1957).

\bibitem{Zerilli} F. Zerilli, Perturbation analysis for gravitational and
electromagnetic radiation in a Reissner-Nordstr\"{o}m geometry, Phys. Rev. D 
\textbf{9}, 860 (1974).

\bibitem{KucharCGW} K. Kucha\v{r}, Canonical Quantization of Cylindrical
Gravitational Waves, Phys. Rev. D \textbf{4}, 955 (1971).

\bibitem{Torre} C. G. Torre, Midi-superspace Models of Canonical Quantum
Gravity, J. Theor. Phys. \textbf{38}, 1081 (1999) [arXiv:gr-qc/9806122].

\bibitem{BojowaldBook} M. Bojowald, \textit{Canonical Gravity and
Applications: Cosmology, Black Holes, and Quantum Gravity} (Cambridge,
England, 2011).
\end{thebibliography}
\end{document}